\documentclass[lettersize,journal]{IEEEtran}
\usepackage[colorlinks,urlcolor=black,linkcolor=black,citecolor=black]{hyperref}
\usepackage{booktabs}
\usepackage{amssymb}
\usepackage[ruled]{algorithm2e}
\usepackage{mathtools}
\usepackage{amsmath,amsfonts}
\usepackage{algorithmic}
\usepackage{array}
\usepackage[caption=true,font=normalsize,labelfont=sf,textfont=sf]{subfig}
\usepackage{color}
\usepackage{mathrsfs} 
\usepackage{orcidlink}
\usepackage{textcomp}
\usepackage{stfloats}
\usepackage{url}
\usepackage{verbatim}
\usepackage{graphicx}
\usepackage{cite}
\begin{document}

\title{The Blind Normalized Stein Variational Gradient Descent-Based Detection for Intelligent Random Access in Cellular IoT }

\author{Xin Zhu~\orcidlink{0000-0002-4741-4193},~\IEEEmembership{Student Member,~IEEE}, and
        Ahmet Enis Cetin~\orcidlink{0000-0002-5607-6587},~\IEEEmembership{Fellow,~IEEE}
        % <-this % stops a space
\thanks{Submitted: Xin Zhu was supported by National Science Foundation (NSF) IDEAL 2217023 and AEC was supported in part by NSF Grant 2229659 and DOE DE-SC0023715. (\textit{Corresponding author: Ahmet Enis Cetin.)}}
\thanks{Xin Zhu and Ahmet Enis Cetin are affiliated with the
Department of Electrical and Computer Engineering, University of Illinois
Chicago, Chicago, IL 60607 USA (email: xzhu61@uic.edu; aecyy@uic.edu). }% <-this % stops a space
% \thanks{Corresponding author: Ahmet Enis Cetin; e-mail: aecyy@uic.edu}
\thanks{Copyright (c) 20xx IEEE. Personal use of this material is permitted. However, permission to use this material for any other purposes must be obtained from the IEEE by sending a request to pubs-permissions@ieee.org}
}

\markboth{IEEE Internet of Things Journal, Vol. 00, No. 0, Month 2020}
{Xin. Zhu \MakeLowercase{\textit{et al.}}: The Blind Normalized Stein Variational Gradient Descent-Based Detection for Intelligent Random Access in Cellular IoT}

\maketitle

\begin{abstract}
The lack of an efficient preamble detection algorithm remains a challenge for solving preamble collision problems in intelligent random access (RA) in the cellular Internet of Things (IoT). To address this problem, we present an early preamble detection scheme based on a maximum likelihood estimation (MLE) model at the first step of the grant-based RA procedure. A novel blind normalized Stein variational gradient descent (SVGD)-based detector is proposed to obtain an approximate solution to the MLE model. First, by exploring the relationship between the Hadamard transform and wavelet packet transform, a new modified Hadamard transform (MHT) is developed to separate high-frequency components from signals using the second-order derivative filter. Next, to eliminate noise and mitigate the vanishing gradients problem in the SVGD-based detectors, the block MHT layer is designed based on the MHT, scaling layer, soft-thresholding layer, inverse MHT and sparsity penalty. Then, the blind normalized SVGD algorithm is derived to perform preamble detection without prior knowledge of noise power and the number of active IoT devices. The experimental results show the proposed block MHT layer outperforms other transform-based methods in terms of computation costs and denoising performance. Furthermore, with the assistance of the block MHT layer, the proposed blind normalized SVGD algorithm achieves a higher preamble detection accuracy and throughput than other state-of-the-art detection methods.

\end{abstract}

\begin{IEEEkeywords}
Grant-based random access, early preamble detection,  Internet of
Things (IoT), Hadamard transform, Stein variational gradient descent. %\LaTeX, paper, template, typesetting.
\end{IEEEkeywords}

\section{Introduction}

\IEEEPARstart{G}{iven} the prevalence of massive machine-type communication (mMTC)~\cite{hsu2023hyper,ma2023model}, the forthcoming cellular Internet of Things (IoT) system is required to accommodate the extensive network of devices~\cite{choi2020throughput}. However, there can be challenges related to the potential preamble collision~\cite{ye2023density,choi2021grant,yin2023learning} issues, particularly when a considerable number of IoT devices attempt to access the network simultaneously. Furthermore, effectively detecting or tackling preamble collision is crucial for the implementation of IoT scenarios~\cite{sarker2023internet}. 
To reduce transmission latency and signaling overhead, a range of strategies, such as grant-based and grant-free random access (RA) schemes, are utilized to mitigate preamble collision problems in cellular IoT.

The currently standardized procedure for device access in IoT networks is grant-based RA (GBRA)~\cite{balci2024fairness,hasan2013random,reddy2021successive,bezerra2018rach,zhou2015low}. It utilizes a communication protocol where IoT devices seek permission prior to transmitting data on the network. In the first stage, every active IoT device randomly chooses a preamble from the pool at each time slot. After that, the base station (BS) transmits the RA responses corresponding to the activated preambles. Next, active IoT devices send message 3 (MSG3), requesting the BS to allocate resources for data transmission. Finally, if the preamble sent in Step 1 is chosen by a single user, the BS will grant the corresponding request and send a contention-resolution message 4 to inform the active IoT device of the available resources. Otherwise, a preamble collision happens. The conflicting IoT devices will not be granted permission and will retry access in the next time slot.

Unlike GBRA, each active IoT device directly transmits the preamble and data to the BS without any permission under the grant-free RA (GFRA) schemes~\cite{liu2024grant,bian2023joint,kang2023noma}. 
Additionally, each user is pre-assigned a dedicated preamble, which can also serve as the user ID. In each time slot, the BS first detects preambles to identify active IoT devices. Subsequently, the BS performs channel estimation based on metadata and decodes data using the acquired channel information.
Compared with GBRA, GFRA alleviates preamble collisions and reduces access latency. However, assigning orthogonal preambles to massive IoT devices is not feasible. Therefore, it is a challenging task to detect active IoT devices under GFRA~\cite{liu2018sparse}. Additionally, GFRA often leads to unreliable transmissions~\cite{liu2023grant}. Hence, to ensure high-rate transmissions alongside both extensive access and reliable connectivity,
we consider the GBRA scheme in this paper. 

\subsection{Early preamble collision detection schemes}
In the GBRA scheme, the preamble collision can only be identified in the fourth step. Therefore, in the second step, the BS still allocates wireless resources to collided IoT devices. It results in a waste of wireless resources. To address this issue, the authors of \cite{jang2016early} introduced an early preamble collision detection (e-PACD) scheme based on the tagged preambles at the first step of GBRA. 
The tagged preambles consist of the preambles and tag Zadoff-Chu (ZC) sequences with different root numbers. The e-PACD scheme determines preamble collision by detecting whether the same preamble contains multiple tags. However, the overuse of root sequences escalates the complexity of detection. Besides, as ZC sequences with varying root indexes lack orthogonality, the frequency of false alarms rises alongside the number of root sequences utilized in correlation operations.
Additionally, a single-root preamble sequence-based early collision detection scheme was proposed for satellite-based IoT~\cite{zhen2021early} at the first step of GBRA. In the scheme, all feasible preambles are generated using the ZC sequence with different cyclic shifts. The intricate design of the cyclic shift offset set improves the efficiency of preamble collision detection at the second step. However, the cyclic shift offsets increase the complexity of the preamble design. Moreover, since the single root ZC sequences are employed, the number of preambles is limited, leading to increasing the probability of preamble collision in a dense user scenario.
% Besides, an effective preamble collision resolution solution is proposed for the large-scale random access process in IoT systems~\cite{zhong2021preamble}. This approach utilizes a new sub-preamble structure composed of shared and dedicated sub-preambles to decrease the probability of preamble collisions. Preamble collisions occur only when both the shared and dedicated sub-preambles collide simultaneously. In cases where only one type of sub-preamble experiences collision, the proposed random access scheme can still work. Moreover, a fixed cluster-based collision resolution random access (FCCRRA)~\cite{he2024fixed} is introduced to alleviate substantial pilot collisions by utilizing variations in received power distribution among distributed access points.

\subsection{Preamble Detection Algorithms}
The appropriate preamble detection algorithms are crucial for detecting preamble collisions in advance. 
% Therefore, the GBRA scheme requires an efficient algorithm to proactively detect preamble collisions. 
In \cite{choi2018mcmc}, a maximum a posteriori (MAP) estimation model was proposed for detecting preambles. A low-complexity Markov Chain Monte Carlo (MCMC) algorithm was employed to sample the MAP model to obtain an approximate solution. However, the MAP estimation needs prior knowledge of the number of active IoT devices, which is unknown in practical communication scenarios. 
The MCMC algorithm also increases estimation errors due to the stochastic nature of sampling. To enhance the accuracy of the MCMC-based scheme, the normalized Stein Variational Gradient Descent (NSVGD) detector~\cite{zhu2024stein} was proposed based on the maximum likelihood estimation (MLE) model. 
% Unlike the MAP model, the MLE model does not require the prior distribution of estimated variables. Additionally, 
The NSVGD detector enhances the accuracy of preamble detection through the deterministic update characteristics of particles. However, to improve robustness in low signal-to-noise ratio (SNR) environments, the NSVGD detector introduces a bias term. This bias term necessitates the BS to know the number of active IoT devices.
In practical communication scenarios, it is challenging to obtain the number of active IoT devices in the cell.
Moreover, the authors of \cite{ni2020index} established a maximum likelihood preamble detection model based on non-Bayesian methods. The non-Bayesian method offers lower computational complexity, enabling rapid preamble detection. Nevertheless, 
the non-Bayesian approach cannot be used to detect preamble collisions. 
Furthermore, deep learning methods have been applied to solve detection problems. In \cite{khan2023preamble}, the convolutional neural network-based approach was employed to detect preambles by extracting features from signals.

\begin{figure}[t]
%\vskip 0.2in
\begin{center}
\centerline{\includegraphics[width=1\linewidth]{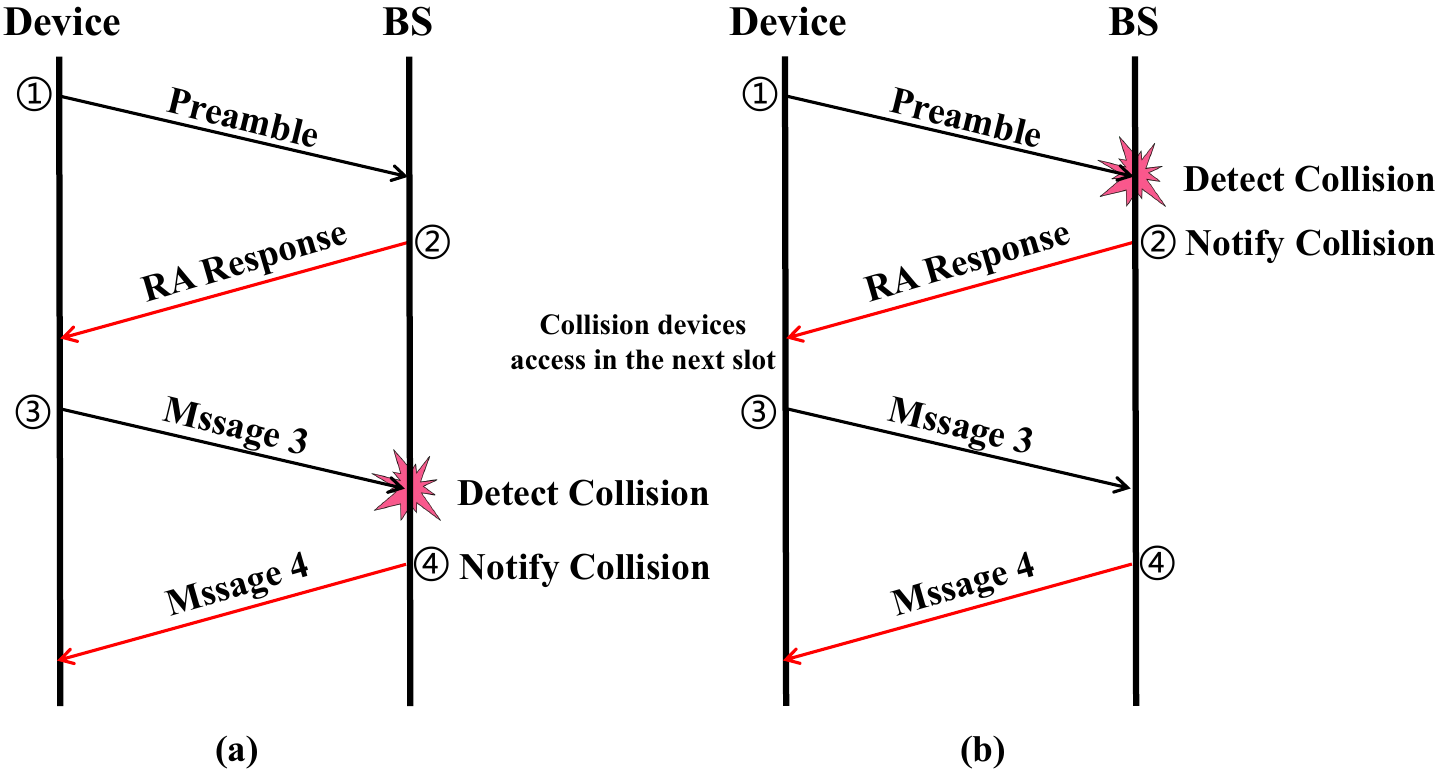}}
\caption{(a) Conventional GBRA. (b) Early preamble collision detection GBRA.}
%\vskip -0.5in
\label{fig: RA}
\end{center}
\end{figure}
\subsection{Motivations and Contributions}
In this paper, we establish an early MLE preamble detection model at the first step of GBRA in cellular IoT. As shown in Fig. \ref{fig: RA}, in contrast to conventional preamble detection, performed in the third step, and conventional collision notification, executed in the fourth step, our model enables rapid preamble detection in the first step and fast collision notification in the second step. Consequently, collided devices do not need to proceed with the third and fourth steps in GBRA, reducing the access latency and waste of resources.

To improve the efficiency of preamble detection, we develop a novel low-complexity blind NSVGD-based detector. More specifically, first, by investigating the connection between the Hadamard transform and wavelet packet transform, we design a new modified Hadamard transform (MHT) by replacing the last half rows in the Hadamard matrix with the second-order derivative filter. Based on the MHT, a new block MHT domain layer is proposed. We developed the discrete cosine transform (DCT) and Hadamard transform domain layers in several applications~\cite{pan2023real,pan2022deep,pan2023hybrid,zhu2024novel}. In this paper, the key idea is to apply the MHT, scaling layer, trainable soft-thresholding layer, and Kullback Leibler divergence-based sparsity penalty to remove the noise from the complex signals. Then, we derive a new blind NSVGD algorithm that combines the SVGD algorithm and momentum strategy to perform preamble detection. The main contributions are concluded as follows: 
\begin{enumerate}

\item A novel MHT is designed by modifying the Hadamard matrix. Compared with the Hadamard transform, the MHT has a better ability to separate high-frequency components from the signals in the transform domain.

\item An efficient block MHT layer is proposed to denoise the signal received by the BS, which helps to alleviate the vanishing gradient problem and enhance the robustness of the NSVGD detector in noisy environments. Compared with other state-of-the-art methods, the block MHT layer achieves a better denoising performance with a lower computation cost.

\item A new blind NSVGD algorithm is derived. In comparison to the NSVGD detector, the blind NSVGD algorithm is capable of finishing preamble detection without the knowledge of noise power and the number of active IoT devices, making it feasible for implementation in practical communication scenarios. 

\item The experimental results show that the proposed preamble detection scheme achieves a higher preamble detection accuracy and better robustness than other baselines under different SNR and different numbers of active IoT devices. Additionally, the proposed method is superior to other schemes in terms of throughput.

\end{enumerate}

The remainder of the paper is structured as follows. Section II describes the maximum likelihood preamble detection model at the first step of GBRA. 
The introduction and error analysis of SVGD-based detectors are provided in Section III. The blind NSVGD-based detector is derived in Section IV.  Simulation results are presented in Section V. Finally, conclusions are given in Section VI.

\textit{Notation}: 
Lowercase letters, bold lowercase letters, and bold uppercase letters represent scalars, vectors, and matrices, respectively.
$(\cdot)^\mathrm{T}$ and $(\cdot)^{\mathrm{H}}$ denotes the transpose and conjugate transpose respectively. 
$(\cdot)_+$ stands for the rectified linear unit (ReLU) function.
$\bigtriangledown(\cdot)$ represents the differentiation operator.
The set of complex numbers is represented by $\mathbb{C}$, while the set of real numbers is indicated by $\mathbb{R}$. 
The notation $\mathcal{CN}(\mathbf a,\mathbf B)$ refers to the circularly symmetric complex Gaussian (CSCG) distribution, which has a mean vector $\mathbf a$ and a covariance matrix $\mathbf B$. Furthermore, Table I provides an overview of the symbols commonly used throughout the paper.

% The notation k·k stands for the Euclidean norm
% and |·| denotes the cardinality of a set. E(·) denotes the operator of
% expectation, and < (·) gives the real part of a complex number.
% ) represents the distribution of a zero-mean circularly
% symmetric complex Gaussian (CSCG) random variable with
% 2
% .

\begin{table}[t]
% \begin{small}
\centering
\caption{\textsc{Glossary of Key Symbols.}}
\label{tab: Symbols}
\renewcommand{\arraystretch}{1.3}
\begin{tabular}{cl}
\toprule
\textbf{Notation} & \textbf{Meaning} \\ \hline
$L$ & Length of preambles \\ \hline
$\delta$ & Noise power \\ \hline
$T$ & Number of receive antennas \\ \hline
$M$ & Number of active devices \\ \hline
$K$ & Number of preambles \\ \hline
$\mathbf z_{(m)}$ & Preamble sequence chosen by the $m$-th active device  \\ \hline
$H_{m,t}$ & Channel coefficient from $t$-th antenna to $m$-th device\\ \hline
$\beta$ & Standard deviation of small-scale fading coefficients \\ \hline
$\mathcal{N}_k$ & Index set of active devices choosing preamble $k$ \\ \hline
$x_{k}$ & Number of active devices choosing preamble $k$ \\ \hline
$n$ & Number of particles \\ \hline
$\mathbf{y}_{t}$ & Received signal at the $t$-th antenna\\ 
\bottomrule
\end{tabular}
% \end{small}
\end{table}

\section{System Model}
\label{sec: System}

In this section, we establish the maximum likelihood preamble detection model at the first step of the GBRA in cellular IoT as shown in Fig. \ref{fig: RA}. 
Suppose the BS is equipped with $T$ antennas, and each active IoT device in the cell is equipped with one antenna. 
Each preamble is randomly chosen by each active IoT device from the pool. The length of each preamble is $L$. Then, the signal received by the BS on the $t$-th antenna can be computed as follows:
\begin{equation}\label{Eq:yph}
		\mathbf{y}_{t}=\sum_{m=1}^{M}\mathbf z_{(m)}H_{m,t}\sqrt{e_{m}}+\mathbf n_{t},
\end{equation}
for $t=1,\dots,T$, where $M$ stands for the number of active IoT devices. 
$\mathbf z_{(m)}$ is the preamble sequence chosen by the $m$-th active IoT device. 
$\mathbf{y}_{t}$ denotes the received signal at the $t$-th antenna.
Additionally, $\mathbf n_{t}\sim \mathcal{CN}(\mathbf{0},\delta\mathbf I)$ represents the background noise and $\delta$ represents the noise power. 
$\mathbf I$ stands for the identity matrix of size $L\times L$. $e_{m}$ denotes the transmission power of the active IoT device. 
$H_{m,t}$ indicates the channel coefficient:
\begin{equation}\label{Eq:channel}
		H_{m,t}=\theta_{m,t}\sqrt{g_{m}},
\end{equation}
where $\theta_{m,t}\sim \mathcal{CN}(0,\beta^2)$ denotes the small-scale fading coefficient between the $t$-th antenna and $m$-th device; $g_{m}$ represents the large-scale fading coefficient.

Assume the number of preambles is $K$. $K> L$. $x_{k}\in[0, M]$ denotes the number of active IoT devices choosing the $k$-th preamble. 
As our target is to detect preamble collisions in the first step of GBRA, we concentrate on estimating $\mathbf x =[x_{1},\dots,x_{k},\dots,x_{K}]^\mathrm{T}$ in the subsequent steps: First of all, the likelihood of $\mathbf x $ is computed as follows:
   \begin{equation}\label{Eq:Lik}
   		{\rm {Lik}(\mathbf x)}=f({\mathbf y_{t}}\mid \mathbf x).
   \end{equation}
% where $f({\mathbf y_{j}}\mid \mathbf x)$ represents the Likelihood function of $\mathbf x$ given for $\mathbf y_{j}$.
After that, let $\mathbf v_{t}=[v_{1,t},v_{2,t},\dots,v_{k,t},\dots,v_{K,t}]^\mathrm{T}$, and $v_{k,t}$ is defined as follows: 
\begin{equation}\label{Eq: vkt}
	v_{k,t}=\sum_{m\in\mathcal{\mathcal{N}}_k}H_{m,t}\sqrt{e_{m}},
\end{equation}
where $\mathcal{N}_k$ represents the index set of active IoT devices choosing the $k$-th preamble. 
Additionally, power control is employed to ensure that signals received at the BS have the same power level. Therefore, $\sqrt{e_m g_m}=1$. Then, according to Eq.~(\ref{Eq:channel}), we have:
\begin{equation}\label{Eq: vktf}
v_{k,t}=\sum_{m\in\mathcal{\mathcal{N}}_k}\theta_{m,t},
\end{equation}
\begin{equation}\label{Eq: vx}
		f(\mathbf v_{t}|\mathbf x)=\prod \limits_{k\in\mathcal{\zeta^{+}(\mathbf x)}} \frac{1}{\pi \beta^2 x_{k}} \exp(-\frac{{\lvert v_{k,t}\rvert}^2}{\beta^2 x_{k}}),
\end{equation}
% Next, suppose the signal transmitted by the active device experiences independent Rayleigh fading:
% \begin{equation}\label{201}
% 		H_{i,j}\sim \mathcal{CN}(0,\delta^2)
% \end{equation}
where $\mathcal{\zeta^{+}(\mathbf x)}=\{k\mid x_k > 0\}$. 
Next, let $\mathbf Z = [\mathbf z_{1}, \dots, \mathbf z_{K}] $. $\mathbf{y}_{t}$ in the Eq.~(\ref{Eq:yph}) can be caculated as follows:
\begin{equation}\label{Eq: yt}
		\mathbf y_{t}=\mathbf Z\mathbf v_{t}+\mathbf n_{t}.
\end{equation}
%for $j=1, \dots, K$.
According to $\mathbf n_{t}\sim \mathcal{CN}(0,\delta\mathbf I)$, we have:
\begin{equation}\label{Eq: yv}
	\begin{array}{l}
		f(\mathbf y_{t}|\mathbf v_{t})=\frac{1}{(\pi \delta)^{L}} \exp(-\frac{1}{\delta}\|\mathbf y_{t}-\mathbf Z\mathbf v_{t}\|^{2}).
	\end{array}
\end{equation}

From Eq. (\ref{Eq: vx}), Eq. (\ref{Eq: yt}) and Eq. (\ref{Eq: yv}), the mean value of $\mathbf y_{t}$ is zero and its covariance matrix is computed as follows:
\begin{equation}\label{Eq: Eyy}
		\mathbb{E}[\mathbf y_{t}\mathbf y_{t}^{\mathrm{H}} \mid \mathbf x]=\beta^2\mathbf Z\mathbf C_{\mathbf x}\mathbf Z^{\mathrm{H}}+\delta\mathbf I=\varphi(\mathbf{x}),
\end{equation}
where $\mathbf C_{\mathbf x}$ = diag$(x_{1}\dots x_{K})$. 
$(\cdot)^{\mathrm{H}}$ represents the conjugate transpose. 
From Eq. (\ref{Eq: Eyy}), $\mathbf y_{t}$ is a CSCG vector:
\begin{equation}\label{Eq: ysimcn}
		\mathbf  y_{t}\mid\mathbf  x\sim \mathcal{CN}(\mathbf 0,\varphi(\mathbf{x})),
\end{equation}
%After that, $\ln f(\mathbf y_{j}\mid\mathbf x)$ can be computed as:
% \begin{equation}\label{Eq:lnfyx}
% \begin{split}
% 		\ln f(\mathbf y_{j}\mid\mathbf x)		
% 		=-\mathbf y_{j}^{\mathrm{H}}(\phi(\mathbf{x}))^{-1}\mathbf y_{j}-\ln\det(\phi(\mathbf{x}))+\xi,
% \end{split}
% \end{equation}
\begin{equation}\label{Eq: fytxe}
	\begin{array}{l}
		f(\mathbf y_{t}\mid\mathbf x)=\frac{1}{(\pi^{L})\det(\varphi(\mathbf{x}))} e^{-\mathbf y_t^{\mathrm{H}}\varphi(\mathbf{x})^{-1}\mathbf y_{t}}.
	\end{array}
\end{equation}
Then $\{\mathbf{y}_t\}_{t=1}^{T}$ is defined as the set of received signals across $T$ antennas. The likelihood function of $\mathbf x$ is:
\begin{equation}\label{Eq: fyxprod}
    f(\{\mathbf{y}_t\}_{t=1}^{T}\mid\mathbf x)=\prod\nolimits_{t=1}^T f(\mathbf y_{t}\mid\mathbf x).
\end{equation}
Furthermore, the log-likelihood function is:
\begin{align}    \label{Eq:lnfyx}
	\begin{aligned} 
    \ln f(\{\mathbf{y}_t\}_{t=1}^{T}\mid\mathbf x)&=\sum\limits_{t=1}^{T}\ln f(\mathbf y_{t}\mid\mathbf x)\\
    &=\sum\limits_{t=1}^{T}-\mathbf y_{t}^{\mathrm{H}}\varphi(\mathbf{x})^{-1}\mathbf y_{t}-\ln\det(\varphi(\mathbf{x}))+\eta,
	\end{aligned}
\end{align}
where $\eta$ is a constant. 
Finally, the maximum log-likelihood estimation (MLE) model can be obtained as follows:
% \begin{align}    
% 	\begin{aligned} 
%     \tilde{\mathbf x}
%     &=\mathop{\arg\max}\sum\limits_{t=1}^{T}-\mathbf y_{t}^{\mathrm{H}}(\omega(\mathbf{x}))^{-1}\mathbf y_{t}-\ln\det(\omega(\mathbf{x}))+\eta
% 	\end{aligned}
% \end{align}
\begin{equation}\label{Eq: argx}
			\tilde{\mathbf x}=\mathop{\arg\max} \ln f(\{\mathbf{y}_t\}_{t=1}^{T}\mid\mathbf x).
\end{equation}
From Eq. (\ref{Eq: argx}), the computational complexity of MLE is $(M+1)^K$, which grows exponentially with $K$. Hence, as $K$ increases, the computational complexity escalates significantly.

\section{svgd based preamble detection}
Due to the high computational complexity associated with directly solving the MLE model, some variational inference methods are applied to obtain an approximate solution, e.g., SVGD-based algorithms. Therefore, we introduce the SVGD-based approaches for the preamble detection problem in this section. 

% In this section, we consider using the SVGD algorithm. 

\subsection{Stein Variational Gradient Descent (SVGD)}

SVGD~\cite{liu2016stein,liu2016kernelized,zhao2023stein,nusken2023geometry} is a particle-based variational inference algorithm used for gradient descent optimization. 
The algorithm iteratively updates particles to gradually transition from the initial distribution to the target distribution by minimizing the KL divergence between the two distributions. A set of particles are first initialized with an arbitrary distribution. Next, 
the particles are updated using the optimal perturbation direction, which corresponds to the steepest descent on the KL divergence. 
After multiple iterations, the particles converge towards the target distribution.
By leveraging the gradient information of the target distribution, SVGD offers a flexible and scalable approach to take samples from the complicated distribution.
% \begin{algorithm}[htbp]
% 	\caption{SVGD}
%         \label{alg: SVGD}
% 	%\LinesNumbered %要求显示行号
% 	\KwIn{A target function $g(\mathbf x)$ and  a set of initial paticles $\{\mathbf x_{u}^0\}_{u=1}^n$}%输入参数
% 	\KwOut{A set of particles $\{\mathbf x_{u}\}_{u=1}^n$ that approximates the target function}%输出
	
% 	\For{iteration $r$}{
% 		$\mathbf x_{u}^{r+1}\leftarrow \mathbf x_{u}^{r}+\varepsilon\varphi(\mathbf x_u^{r})$ where $\varphi(\mathbf x)=\frac{1}{n}\sum_{j=1}^{n}[k(\mathbf x_j^{t},\mathbf x)\bigtriangledown_{\mathbf x_j^{t}}\log g(\mathbf x_j^{t})+\bigtriangledown_{\mathbf x_j^{t}}k(\mathbf x_j^{t},\mathbf x)]$
		
% 		where $\varepsilon$ is the step size. $k(x,\cdot)$ is kernel function.  			
% 	}
% \end{algorithm}

\subsection{SVGD Detector}
To solve the preamble detection problem, we employed the SVGD algorithm to obtain an approximate solution to the MLE model \cite{zhu2024stein}. The key idea is applying SVGD to sample $\mathbf x$ from the target distribution $p(\mathbf x)$. Here, the density function of $p(\mathbf x)$ is $g(\mathbf x)=f(\{\mathbf y_{t}\}\mid\mathbf x)$.
It is mentioned in \cite{liu2016stein} that the particles can be initialized with any distribution. 
Therefore, we first use the uniform distribution to initialize a set of particles $\{\mathbf x_{u}\}_{u=1}^n$. $n$ is the number of particles.
Then, for $r$-th iteration, the particles are updated as follows: 
\begin{equation}\label{Eq: iter}
		\mathbf x_{u}^{r+1}\leftarrow \mathbf x_{u}^{r}+\lambda{\omega}(\mathbf x_u^{r}),
\end{equation}
where $\lambda$ is a step size. $\omega(\cdot)$ is a velocity field, which transports the particles to approximate the target distribution. It is defined as: 
 \begin{equation}\label{Eq:gradient}
 {\omega}(\mathbf x)=\frac{1}{n}\sum_{i=1}^{n}[k(\mathbf x_i^{r},\mathbf x)\bigtriangledown_{\mathbf x_i^{r}}\ln g(\mathbf x_i^{r})+\bigtriangledown_{\mathbf x_i^{r}}k(\mathbf x_i^{r},\mathbf x)],
 \end{equation}
where $k(\mathbf x_i^{r},\mathbf x)=\exp(-\frac{1}{h}||\mathbf x_i^{r}-\mathbf x||_{2}^2)$ is the Gaussian radial basis function (RBF) kernel function~\cite{kuo2013kernel}. $h={\rm{med}}^2/\ln n$. ${\rm{med}}$ is the median of pairwise distances between $\{\mathbf {x}_{u}\}_{u=1}^n$. After sufficient iterations, the particles $\{\mathbf {\tilde{x}}_{u}\}_{u=1}^n$ sampled by SVGD approximate the target distribution $p(\mathbf x)$. Next, $\{\mathbf {\tilde{x}}_{u}\}_{u=1}^n$ is utilized to estimate the number of IoT devices choosing each preamble.

\subsection{NSVGD Detector}
\label{sec: NSVGD}
% However, the detection performance of SVGD will be affected by environmental noise.
From Eq.~(\ref{Eq:lnfyx}), it is observed that the value of the log-likelihood function is mainly determined by $\varphi(\mathbf{x})=\beta^2\mathbf Z\mathbf C_{\mathbf x}\mathbf Z^{\mathrm{H}}+\delta\mathbf I$. $\delta$ is noise power.
When the modulus of each entry $\tau$ in matrix $\phi(\mathbf{x})=\beta^2\mathbf Z\mathbf C_{\mathbf x}\mathbf Z^{\mathrm{H}}$ is much smaller than $\delta$, we have $\varphi(\mathbf{x}) \thickapprox \delta\mathbf I$, which indicates 
$\varphi(\mathbf{x})$ is independent to the change of $\phi(\mathbf{x})$. Furthermore, Eq.~(\ref{Eq:lnfyx}) is rewritten as:
\begin{equation}\label{Eq: lngx}
		\ln g(\mathbf x_i^{r})
		=-\sum_{t=1}^{T}\mathbf y_t^{\mathrm{H}}\mathbf (\delta\mathbf I)^{-1}\mathbf y_{t}-T\ln\det(\delta\mathbf I)+\eta.	
\end{equation}
Hence, $\bigtriangledown_{\mathbf{x}_u^{r}}\ln g(\mathbf x_u^{r})=0$. Then ${\omega}(\mathbf x)$ is caculated as:
\begin{equation}\label{Eq:GL}
		{\omega}(\mathbf x)=\frac{1}{n}\sum_{i=1}^{n}[\bigtriangledown_{\mathbf x_i^{r}}k(\mathbf x_i^{r},\mathbf x)].
\end{equation}
According to Eq.~(\ref{Eq:GL}), it is noticed that only $\bigtriangledown_{\mathbf x_i^{r}}k(\mathbf x_i^{r},\mathbf x)$ is applied to update $\mathbf x$. However, $\bigtriangledown_{\mathbf x_i^{r}}k(\mathbf x_i^{r},\mathbf x)$ does not contain any effective information about the target distribution. Then, the particles are updated towards the wrong direction. Therefore, the noise can cause the vanishing gradients problem in the SVGD detector, which increases the estimation errors. To enhance the robustness of the SVGD-based model, we further proposed NSVGD based on the momentum and a bias correction term $\vartheta$~\cite{zhu2024stein}, which is defined as:
\begin{equation}\label{Eq:error}
    \vartheta = \nu \left(nM-\sum_{u=1}^{n}\vert\vert\mathbf{x}_{u}^{r}\vert\vert_{1}\right),
\end{equation}
where $\nu$ is constant weight. $\vartheta$ utilizes the number of active IoT devices in the cell as a constraint to correct the direction of particle updates. However, the number of active IoT devices is unknown in practical communication scenarios. Therefore, the NSVGD detector cannot directly be applied in the random access scheme. 

\section{The blind NSVGD-based detector for preamble detection}
%\subsection{The modified Hadamard aided NSVGD detector}
As described above, the performance of the SVGD detector is susceptible to environmental noise. Additionally, the NSVGD detector~\cite{zhu2024stein} requires unknown prior knowledge to detect the preamble. To address these issues, we propose a novel blind NSVGD-based detector. Firstly, a new modified Hadamard transform (MHT) is designed to separate
high-frequency components from the signals utilizing the second-order derivative filter. Next, the block MHT layer is developed to eliminate noise using the MHT, scaling layer, trainable soft-thresholding layer and inverse MHT. Then, a new blind NSVGD algorithm is designed for preamble detection without requiring unknown prior knowledge.

\subsection{The modified Hadamard transform (MHT)}
% Since the SVGD or NSVGD-based detector can not extract valid information from the target distribution under low SNR, 
Transform-based denoising methods have become prevalent in data analysis for efficiently reducing noise from signals. These techniques apply mathematical transforms to represent signals in alternate domains, enabling the separation of noise from the underlying structure. Commonly used transforms include the Fourier transform, wavelet transform, Hadamard transform, and discrete cosine transform (DCT). Among them, the Hadamard transform concentrates the majority of the signal energy in a small subset of coefficients. The rest of the coefficients are redundant data. Such energy concentration property allows for separating noise from important components in the transform domain, thus enhancing denoising effectiveness.
Given an input vector $\mathbf{x}=[x_0,x_1,\dots,x_{D-1}]$, its Hadamard transform $\mathbf{X}=[X_0,X_1,\dots,X_{D-1}]$ is defined as:
% \begin{equation}
%     {X}_k=\sqrt{\frac{1}{N}}\sum_{n=0}^{N-1}x_n (-1)^{\sum_{i=0}^{M-1}k_{i}n_{i}},
% \label{Eq:dct3}
% \end{equation}
% where $k_i$ and $m_i$ are the $i$-bits in the binary representations
% of $k$ and $m$, respectively. 
\begin{equation}
    \mathbf{X}=\mathcal{T}(\mathbf{x})=\sqrt\frac{1}{D} \mathbf{H}_D\mathbf{x},
\end{equation}
where the Hadamard matrix $\mathbf{H}_D$ is computed as follows:
\begin{equation}\label{eq: Hadamard matrix}
			\mathbf{H}_D = 
			\begin{cases}
				1,& D = 1,\\
				\begin{pmatrix*}[r]
					1 & 1 \\ 1 & -1
				\end{pmatrix*},& D =2,\\
				\begin{pmatrix*}[r]
					\mathbf{H}_{\frac{D}{2}} & \mathbf{H}_{\frac{D}{2}} \\ \mathbf{H}_{\frac{D}{2}} & -\mathbf{H}_{\frac{D}{2}}
				\end{pmatrix*},& D \ge 4.
			\end{cases}
		\end{equation}
		% Alternatively, for $N\ge4$, $N=2^M$, $\mathbf{H}_N$ can also be computed using Kronecker product $\otimes$:
		% \begin{equation}
		% 	\mathbf{H}_{N}=\mathbf{H}_2 \otimes\mathbf{H}_{\frac{N}{2}}=\mathbf{H}_2^{\otimes M}.
		% \end{equation}

The inverse Hadamard transform is defined as:
\begin{equation}
    \mathbf{x}=\mathcal{T}^{-1}(\mathbf{X})=\sqrt\frac{1}{D} \mathbf{H}_D\mathbf{X}=\mathcal{T}(\mathbf{X}).
\end{equation}

\begin{figure}[t]
%\vskip 0.2in
%\begin{center}
\centerline{\includegraphics[width=0.8\linewidth]{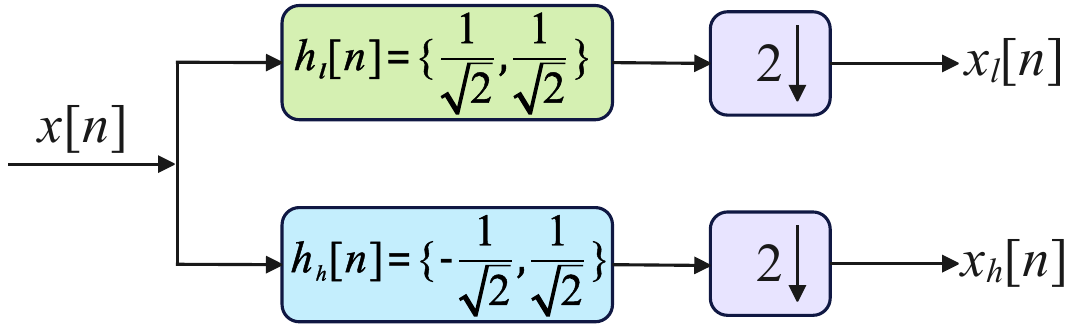}}
\caption{The two-channel decomposition filter bank.}
%\vskip -0.5in
\label{fig: two-channel}
%\end{center}
\end{figure}

\begin{figure}[t]
%\vskip 0.2in
\begin{center}
\centerline{\includegraphics[width=1\linewidth]{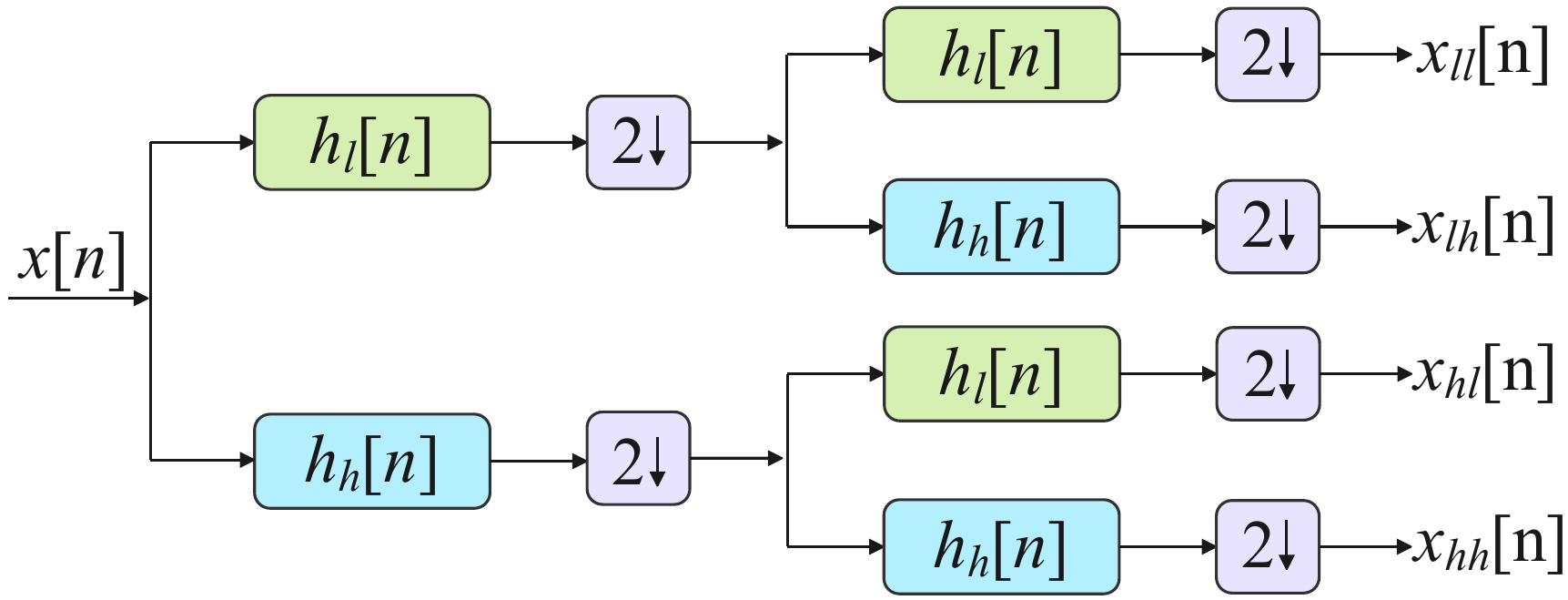}}
\caption{The four-channel decomposition filter bank.}
%\vskip -0.5in
\label{fig: four-channel}
\end{center}
\end{figure}

Another interpretation of the Hadamard transform is related to the wavelet packet transform~\cite{cetin1993block}. The Hadamard transform can be constructed using a decomposition filter bank. The two-channel decomposition filter bank has a low-pass filter $h_l[n]=\{\frac{1}{\sqrt{2}},\frac{1}{\sqrt{2}}\}$
and the high-pass filter $h_h[n]=\{-\frac{1}{\sqrt{2}},\frac{1}{\sqrt{2}}\}$ respectively, as shown in Fig. \ref{fig: two-channel}.
Assume that $x[n]=\{\cdots,x_0,x_1,x_2,x_3,\cdots\}$ is peiodic extension of ${x_0, x_1}$. In this case, $x_l[n]=\{\cdots,\frac{x_0+x_1}{\sqrt{2}},\frac{x_0+x_1}{\sqrt{2}},\cdots\}$ and $x_h[n]=\{\cdots,\frac{x_0-x_1}{\sqrt{2}},\frac{x_0-x_1}{\sqrt{2}},\cdots\}$, respectively. Therefore, 
\begin{equation}\label{eq: Hadamard matrix2}
				\begin{pmatrix*}[r]
					x_l \\ x_h
				\end{pmatrix*}\\ = \frac{1}{\sqrt{2}}
				\begin{pmatrix*}[r]
					1 & 1 \\ 1 & -1
				\end{pmatrix*}
    				\begin{pmatrix*}[r]
					x_0 \\ x_1
				\end{pmatrix*},
    \\
		\end{equation}
where the input to the filter bank is related to the output via the two-by-two Hadamard transform matrix. Next, consider the wavelet packet transform shown in Fig. \ref{fig: four-channel}.

Let $x[n]=\{\cdots,x_0,x_1,x_2,x_3,x_0,x_1,x_2,x_3,\cdots\}$ is a periodic extension of $\{x_0, x_1,x_2, x_3\}$. We have 
\begin{align}    
	\begin{aligned} 
		&x_{ll}[n]=\{\cdots,\frac{x_0+x_1+x_2+x_3}{2},\frac{x_0+x_1+x_2+x_3}{2},\cdots\},\\
		&x_{lh}[n]=\{\cdots,\frac{x_0+x_1-x_2-x_3}{2},\frac{x_0+x_1-x_2-x_3}{2},\cdots\},\\
  	&x_{hl}[n]=\{\cdots,\frac{x_0-x_1+x_2-x_3} 
        {2},\frac{x_0-x_1+x_2-x_3}{2},\cdots\},\\
		&x_{hh}[n]=\{\cdots,\frac{x_0-x_1-x_2+x_3}{2},\frac{x_0-x_1-x_2+x_3}{2},\cdots\}.\\
	\end{aligned}
 \nonumber 
\end{align}
Therefore,
\begin{equation}\label{eq: Hadamard matrix4}
				\begin{pmatrix*}[r]
					x_{ll}[0] \\ x_{lh}[0] \\x_{hl}[0] \\x_{hh}[0]
				\end{pmatrix*}\\ = \frac{1}{2}
				\begin{pmatrix*}[r]
					1 & 1 & 1 & 1 \\
                    1 & 1 & -1 & -1 \\
                    1 & -1 & 1 & -1 \\
                    1 & -1 & -1 & 1 \\
				\end{pmatrix*}
    				\begin{pmatrix*}[r]
					x_{0} \\ x_{1} \\x_{2} \\x_{3}
				\end{pmatrix*},
    \\
		\end{equation}
where the matrix establishing the relationship between input and output has the same rows as the $4\times4$ Hadamard transform. Therefore, the last two rows in the Hadamard matrix can be considered as ``high-pass filters'' to extract the high-frequency components, e.g., noise.
Similarly, $8\times8$ Hadamard transform can be constructed from a $\log_2 8=3$ layers two-channel filter bank. Then, the last four rows in the Hadamard matrix can be considered as ``high-pass filters''.
The concept can be generalised to $N$ by $N$ Hadamard transform which can be constructed using a basic two-channel Haar filter bank. Moreover, it is observed that the Hadamard transform requires the size of the input to be a power of 2. However, the signal received by the BS can be of arbitrary size. To solve this problem, we divide $\mathbf y_{t}$ into short-time windows of length 8. If the size of the signal is not a multiple of 8, zeros will be padded at the end of the signal.

Although the last half row vectors in the Hadamard matrix work as high-pass filters, these row vectors may not be the optimal choice. Inspired by this, we expect to use more efficient high-pass filters to replace the last half rows in the Hadamard matrix. At present, the derivative operator has been widely used as a high-pass filter~\cite{kim1992class,mlsna2009gradient}. In the case of RA, the derivative of $\mathbf{y}_{t}[i]$ is defined as follows:
\begin{equation}\label{Eq:y+h}
	 \mathbf{y}_{t}'[i]=\lim_{\Delta h\rightarrow0}\frac{\mathbf{y}_{t}[i+\Delta h]-\mathbf{y}_{t}[i]}{\Delta h}.
\end{equation}
For a discrete signal, the smallest interval $\Delta h$ is 1. Therefore, the derivative is calculated as $\mathbf{y}_{t}[i+1]-\mathbf{y}_{t}[i]$. Similarly, we have $\mathbf{y}_{t}'[i+1]=\mathbf{y}_{t}[i+2]-\mathbf{y}_{t}[i+1]$. Furthermore, the second derivative of $\mathbf{y}_{t}[i]$ can be calculated as:
\begin{align}    
            \mathbf{y}_{t}''[i]&=\mathbf{y}_{t}'[i+1]-\mathbf{y}_{t}'[i]\nonumber\\
            &=\mathbf{y}_{t}[i+2]-\mathbf{y}_{t}[i+1]-\mathbf{y}_{t}[i+1]+\mathbf{y}_{t}[i]\nonumber\\
            &=\mathbf{y}_{t}[i+2]-2\mathbf{y}_{t}[i+1]+\mathbf{y}_{t}[i].
\end{align}
It is observed that calculating the second-order derivative is identical to performing a convolution operation with the filter [1,-2,1].
The second-order derivative filter is widely utilized in various wavelet transforms. It is the basis of wavelet from the hat function (Section 1.2 in~\cite{strang1989wavelets}).
Additionally, by taking the discrete-time Fourier transform (DTFT)~\cite{sundararajan2024discrete} of $\mathbf{y}_{t}''[i]$, the frequency response of the second-order derivative filter is obtained as $H_{t}(w)=e^{2jw}-2e^{jw}+1=-4\sin^{2}{\frac{w}{2}}e^{jw}$, where $w\in [0, \pi]$ denotes the frequency. 
Then, the magnitude of frequency response $|H_{t}(w)|=4\sin^{2}{\frac{w}{2}}$.
It is observed that $|H_{t}(w)|$ increases as 
$w$ increases for $0 \le w \le \pi$, indicating that the second-order derivative functions as a high-pass filter. This property enables its application in MHT to isolate high-frequency components, such as noise, from the original signal. Hence, we replace the last half rows in the Hadamard matrix with the second derivative filter [1,-2,1]. 
Besides, we shift the filter across different rows to high-pass filter the input vectors as much as possible. In terms of the $8\times8$ Hadamard transform, we modify the Hadamard matrix as follows:
\begin{equation}
\mathbf{{Q}_8}=
\begin{pmatrix*}[r]
     1&  1&  1&  1&  1&  1&  1&  1\\
     1& -1&  1& -1&  1&  -1&  1&  -1\\
     1&  1& -1& -1&  1&  1&  -1&  -1\\
     1& -1& -1&  1&  1&  -1& -1&  1\\
     1&  -2&  1&  0&  0&  0&  0& 0\\
     0&  0&  1& -2&  1&  0&  0&  0\\
     0&  0&   0&  1&  -2&  1&  0&  0\\
     0&  0&   0&  0&  0&  1&  -2&  1\\

    % 1 & 1 \\ 1 & -1
     % 1&  1&  1&  1&  1&  0&  0&  0\\
     % 1& -1&  1& -1& -2&  0&  0&  0\\
     % 1&  1& -1& -1&  1&  1&  0&  0\\
     % 1& -1& -1&  1&  0&  -2& 1&  0\\
     % 1&  1&  1&  1&  0&  1&  -2& 0\\
     % 1& -1&  1& -1&  0&  0&  1&  1\\
     % 1&  1& -1& -1&  0&  0&  0&  -2\\
     % 1& -1& -1&  1&  0&  0&  0&  1\\
\end{pmatrix*}.
\end{equation}
After that, for a block input $\mathbf{x}\in\mathbb{R}^{8}$, the MHT is defined as $\mathbf{y}=\mathbf{{Q}_8}\mathbf{x}$. The same structure can be extended to higher dimensions in a straightforward manner.
In this paper, we only use $8\times8$ MHT because a small block size is very helpful in reducing the computation cost and the number of trainable parameters which will be introduced in section \ref{sec: The block MHT layer}. Furthermore, the small-sized MHT reduces latency due to its low computation cost, especially in a massive RA procedure.
Additionally, it is experimentally shown in section \ref{sec: simulation} that the MHT has a better capability in separating noise from the main components in the frequency domain compared with the Hadamard transform.
% Furthermore, there are four multiplication operations in the modified Hadamard transform, which increase the Multiply–Accumulates (MACs). Therefore, We split the multiplication operation into two addition operations, for example, $-2\times \mathbf{y}_{t}[i]=-\mathbf{y}_{t}[i]-\mathbf{y}_{t}[i]$. Compared with standard Hadamard matrix, the modified Hadamard matrix includes more zeros, which reduces the number of additions from 63 to 47 for each block of received signals. Their numbers of multiplications are both zero.

\subsection{The block MHT layer}
\label{sec: The block MHT layer}

\begin{figure*}[t]
%\vskip 0.2in
\begin{center}
\centerline{\includegraphics[width=1\linewidth]{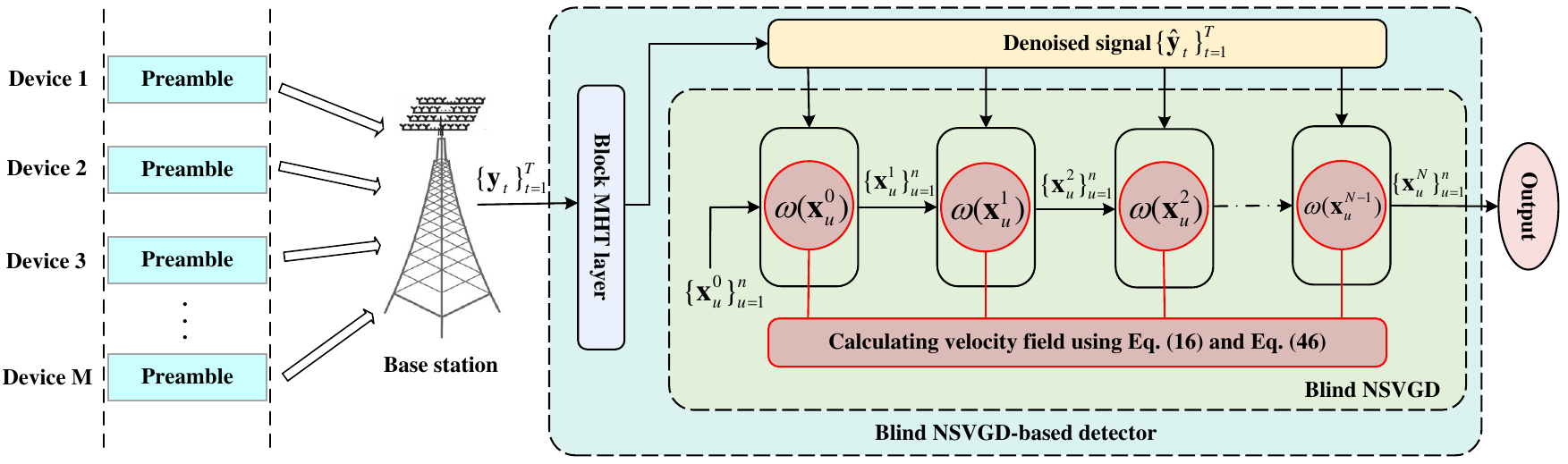}}
\caption{Flowchart of the preamble detection procedure.}
%\vskip -0.5in
\label{fig: Flowchart}
\end{center}
\end{figure*}

\begin{figure}[t]
%\vskip 0.2in
\begin{center}
\centerline{\includegraphics[width=1\linewidth]{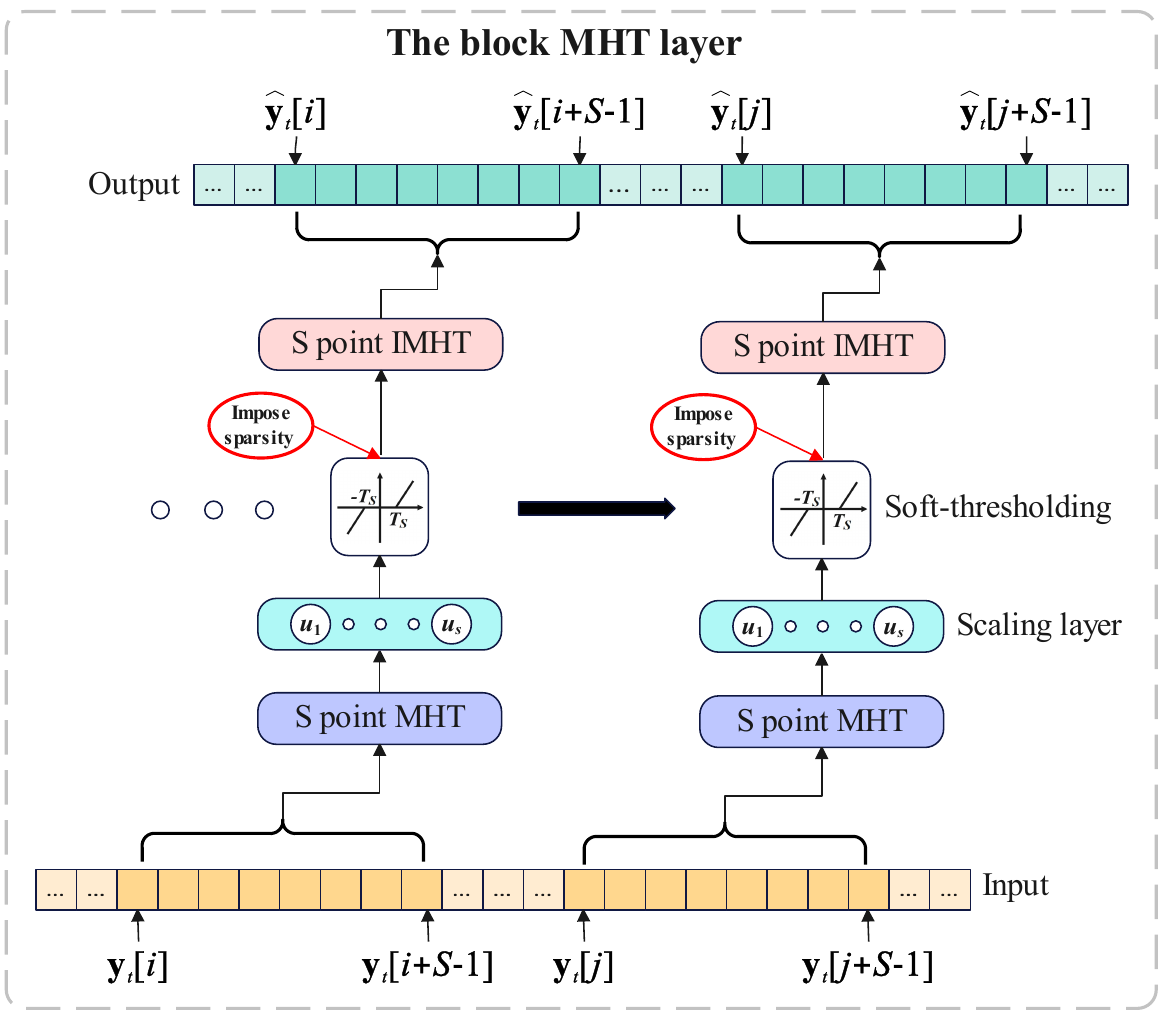}}
\caption{The block MHT layer.}
%\vskip -0.5in
\label{fig: MHT}
\end{center}
\end{figure}

As shown in Fig. \ref{fig: Flowchart}, the block MHT (BMHT) layer is designed to remove the noise from the complex signals $\mathbf{y}_{t}$. 

\subsubsection{Data preprocessing}
we consider the received signal $\mathbf y_t\in\mathbb{C}^{L}$ on a single antenna as one data sample. 
Since neural networks cannot be trained with complex numbers, we concatenate the real and imaginary parts of $\mathbf y_t$ as $\mathbf y_t^c\in\mathbb{R}^{2L}$. Next, we divide $\mathbf y_t^c$ into the small blocks with a size of $S=8$.

\subsubsection{Structure of the BMHT layer}
as shown in Fig. \ref{fig: MHT}, we first perform the MHT on each input block $\mathbf{\tilde{y}}_t\in\mathbb{R}^{S}$. It is computed as:
\begin{equation}\label{Eq: MHT}
    \mathbf{\mathring{y}}_{t}=\mathscr{H}(\mathbf{\tilde{y}}_{t})={Q}_S\mathbf{\tilde{y}}_{t}.
\end{equation}
Then, we employ the scaling layer to allocate weights for different frequency components. It is defined as:
\begin{eqnarray}\label{Eq: scaling}
\mathbf{\bar{y}}_{t}=\mathbf{\mathring{y}}_{t}\circ \mathbf{u},
\end{eqnarray}
where $\circ$ represents the element-wise multiplication. $\mathbf{u}\in\mathbb{R}^{S}$ is the scaling vector, which is trained using the back-propagation algorithm~\cite{cilimkovic2015neural}. 

 After that, the trainable soft-thresholding layer is utilized to eliminate small entries or scale the large entries in the modified Hadamard domain. The small entries are usually noise and redundant information. The soft-thresholding function is defined as:
\begin{equation}\label{Eq: soft}
    \mathbf{p}_t=\mathcal{S}_{T}\left(\mathbf{\bar{y}}_{t}\right)= \text{sign}\left(\mathbf{\bar{y}}_{t}\right)\cdot\left(|\mathbf{\bar{y}}_{t}|-\mathbf{T}\right)_{+},
\end{equation}
where $\mathbf{T}\in \mathbb{R}^{S}$ is the threshold vector trained using the back-propagation algorithm; After the soft-thresholding layer, we perform the inverse modified Hadamard transform: 
\begin{equation}\label{Eq: IMHT}
    \mathbf{\hat{y}}_t=\mathscr{H}^{-1}(\mathbf{p}_t)=Q_S^{-1}\mathbf{p}_t,
\end{equation} 
where 
\begin{equation}
Q_S^{-1}=
\begin{pmatrix*}[r]
    % 1 & 1 \\ 1 & -1
     \frac{1}{8}&\frac{5}{8}&\frac{1}{8}&\frac{5}{8}&-\frac{1}{2}&\frac{1}{2}&{1}&{0}\\[1.5pt]
     \frac{1}{8}&\frac{7}{8}&-\frac{1}{8}&\frac{1}{8}&-{1}&{0}&\frac{1}{2}&\frac{1}{2}\\[1.5pt]
     \frac{1}{8}&\frac{9}{8}&-\frac{3}{8}&-\frac{3}{8}&-\frac{1}{2}&-\frac{1}{2}&{0}&{1}\\[1.5pt]
     \frac{1}{8}&\frac{3}{8}&-\frac{1}{8}&-\frac{3}{8}&{0}&-{1}&-\frac{1}{2}&\frac{1}{2}\\[1.5pt]
     \frac{1}{8}&-\frac{3}{8}&\frac{1}{8}&-\frac{3}{8}&\frac{1}{2}&-\frac{1}{2}&{-1}&{0}\\[1.5pt]
     \frac{1}{8}&-\frac{9}{8}&\frac{3}{8}&-\frac{3}{8}&{1}&{0}&-\frac{1}{2}&-\frac{1}{2}\\[1.5pt]
     \frac{1}{8}&-\frac{7}{8}&\frac{1}{8}&\frac{1}{8}&\frac{1}{2}&\frac{1}{2}&{0}&{-1}\\[1.5pt]
     \frac{1}{8}&-\frac{5}{8}&-\frac{1}{8}&\frac{5}{8}&{0}&{1}&\frac{1}{2}&-\frac{1}{2}\\
\end{pmatrix*}.
\end{equation}

Next, we process the signals block by block. For different blocks, we utilize the same scaling parameters and soft thresholds. Finally, we resize the signal back to its original size by removing the padded zeros.

\subsubsection{ The training of the BMHT layer}:
\label{sec:Sparsity Penalty}
To eliminate noise and redundant information efficiently, we impose sparsity in the modified Hadamard domain during the training stage. Assume the output of the soft-thresholding layer is $\mathbf{p}_t\in \mathbb{R}^S$. Then the activity of ${p}_{t,j}$ is computed as:
\begin{equation}
    \hat{p}_{t,j}={\sigma}(\mathbf{p}_t)_{j}=\frac{e^{ p_{t,j}}}{e^{ p_{t,j}}+1}, j=0,1,\cdots,S-1,
\end{equation}
where $\sigma(\cdot)$ represents the sigmoid function.
Furthermore, the Kullback–Leibler divergence (KLD)~\cite{ng2011sparse} has the capability to measure the distinction between the different distributions. Therefore, we employ KLD as the sparsity penalty term:
\begin{equation}
\sum_{j=0}^{S-1} {\rm{KL}}(\kappa\vert\vert\hat{p}_{t,j})
=\sum_{j=0}^{S-1}\kappa\log\frac{\kappa}{\hat{p}_{t,j}}+
    (1-\kappa)\log\frac{1-\kappa}{1-\hat{p}_{t,j}},
    \label{Eq:KL}
\end{equation}
where $\kappa$ is a sparsity parameter. Hence, for each block, the overall loss function $\mathcal{L}$ of the MHT layer is :
\begin{align}
&\mathcal{L}=\frac{1}{S}\sum_{i=0}^{S-1} \left({\hat{y}}_{t,i}-{\tilde{p}}_{t,i}\right)^2+\rho\sum_{j=0}^{S-1} {\rm{KL}}\left(\kappa\vert\vert{\sigma}(\mathbf{p}_t)_j\right),
    \label{Eq:loss_p}
\end{align}
where $\rho$ is the sparsity penalty weight. ${\tilde{p}}_{t,i}$ and ${\hat{y}}_{t,i}$ stand for the clean and denoised signal on the $t$-th antenna, respectively. Finally, the optimal scaling parameters and soft thresholds are obtained by minimizing the loss function defined in Eq. (\ref{Eq:loss_p}).

\subsection{Blind NSVGD algorithm}
\label{sec: BSVGD}

After the trained BMHT layer denoises the signals, the blind NSVGD algorithm is utilized to perform preamble detection as shown in Fig. \ref{fig: Flowchart}. As mentioned in section \ref{sec: NSVGD}, the NSVGD detector requires prior knowledge of the noise power $\delta$ and the number of active IoT devices. However, this prior knowledge is hard to obtain in practical
communication scenario. 
Therefore, this section presents a new blind NSVGD algorithm capable of performing preamble estimation tasks without unknown prior knowledge. 
As is shown in Algorithm \ref{alg: optimized SVGD}, we first initialize the particles $\{\mathbf x_{u}\}_{u=1}^n$ with a uniform distribution. Next, we start computing $\bigtriangledown_{\mathbf x^{r}}\ln g(\mathbf x^{r})$.  
Since $\delta$ is unknown, we remove term $\delta\mathbf I$ from 
$\varphi(\mathbf{x})$ directly. It is feasible because we have performed denoising previously.
% Therefore, $\varphi(\mathbf{x})=\beta^2\mathbf Z\mathbf C_{\mathbf x}\mathbf Z^{\mathrm{H}}$. 
Then Eq. (\ref{Eq:lnfyx}) is rewritten as:
\begin{align}    \label{Eq:relnfyx}
  \ln g(\mathbf x)= \ln f(\{\mathbf{\hat{y}}_{t}\}\mid\mathbf x)=&\sum\limits_{t=1}^{T}-\mathbf{\hat{y}}_{t}^{\mathrm{H}}(\beta^2\mathbf Z\mathbf C_{\mathbf x}\mathbf Z^{\mathrm{H}})^{-1}\mathbf{\hat{y}}_{t}+\eta\nonumber\\ &-T\ln\det(\beta^2\mathbf Z\mathbf C_{\mathbf x}\mathbf Z^{\mathrm{H}}).
\end{align}
Next, $\bigtriangledown_{\mathbf x^{r}}\ln g(\mathbf x^{r})$ is obtained by computing $\bigtriangledown_{ x_k^{r}}\ln g( \mathbf x^{r})$:
\begin{eqnarray}\label{Eq: dxrlngxr}
	\begin{array}{l}
		\bigtriangledown_{\mathbf x^{r}} \ln g(\mathbf x^{r})=[\bigtriangledown_{x_{1}^{r}}\ln g(\mathbf x^{r}),\dots,\bigtriangledown_{x_{K}^{r}} \ln g(\mathbf x^{r})].
	\end{array}
\end{eqnarray}

From  Eq. (\ref{Eq:relnfyx}), $\bigtriangledown_{ x_k^{r}}\ln g( \mathbf x^{r})$ is computed as follows:
% However, it is challenging to compute $\bigtriangledown_{\mathbf x_l^{t}}\log q(\mathbf x_l^{t})$ directly due to the complex form of $\log q(\mathbf x_l^{t})$. Therefore, we utilize matrix derivative theorems and the chain rule to derive an expression for $\bigtriangledown_{\mathbf x_l^{t}}\log q(\mathbf x_l^{t})$.
% From Eq.(16), we have:
\begin{align}
\bigtriangledown_{ x_k^{r}}\ln g( \mathbf x^{r})=&-\bigtriangledown_{ x_k^{r}}(\sum\limits_{t=1}^{T}\mathbf {\hat{y}}_{t}^{\mathrm{H}}(\beta^2\mathbf Z\mathbf C_{\mathbf x}\mathbf Z^{\mathrm{H}})^{-1}\mathbf{\hat{y}}_{t})\nonumber\\ &-\bigtriangledown_{ x_k^{r}}(T\ln\det(\beta^2\mathbf Z\mathbf C_{\mathbf x}\mathbf Z^{\mathrm{H}})).
\end{align}

Let $\varepsilon(\mathbf{x})=\beta^2\mathbf Z\mathbf C_{\mathbf x}\mathbf Z^{\mathrm{H}}$, we have
\begin{eqnarray}\label{Eq: alldx}
	\begin{array}{l}
		\bigtriangledown_{ x_k^{r}}(\sum\limits_{t=1}^{T}\mathbf {\hat{y}}_{t}^{\mathrm{H}}\varepsilon(\mathbf{x})^{-1}\mathbf {\hat{y}}_{t})=\sum\limits_{t=1}^{T}\mathbf {\hat{y}}_{t}^{\mathrm{H}}\bigtriangledown_{ x_k^{r}}(\varepsilon(\mathbf{x})^{-1})\mathbf{\hat{y}}_{t}.    
	\end{array}
\end{eqnarray}

From inverse matrix derivative lemma~\cite{petersen2008matrix}, $\bigtriangledown_{ x_k^{r}}(\varepsilon(\mathbf{x})^{-1})$ is computed as:
\begin{eqnarray}\label{Eq: inversedx}
\bigtriangledown_{ x_k^{r}}(\varepsilon(\mathbf{x})^{-1})=-\varepsilon(\mathbf{x})^{-1}\bigtriangledown_{ x_k^{r}}(\varepsilon(\mathbf{x}))\varepsilon(\mathbf{x})^{-1},
\end{eqnarray}
% due to the inverse matrix differentiation lemma. Furthermore,
where
\begin{eqnarray}\label{Eq: dx}
	\begin{array}{l}
		\bigtriangledown_{ x_k^{r}}(\varepsilon(\mathbf{x}))=\beta^2 \mathbf z_{k} \mathbf z_{k}^{\mathrm{H}}.
	\end{array}
\end{eqnarray}

From Eq. (\ref{Eq: inversedx}) and Eq. (\ref{Eq: dx}), Eq. (\ref{Eq: alldx}) is rewritten as:
% \begin{eqnarray}\label{CCC}
% 	\begin{array}{l}
% 		\bigtriangledown_{x_{l}^{\ell}}\mathbf V(\boldsymbol x^{\ell})=-\sigma_{\mathrm{H}}^2 \mathbf V(\boldsymbol x^{\ell}) \boldsymbol c_{l} \boldsymbol c_{l}^{\mathrm{H}} \mathbf V(\boldsymbol x^{\ell})
% 	\end{array}
% \end{eqnarray}
% Therefore,
\begin{align}    
	\begin{aligned}        
			\bigtriangledown_{ x_k^{r}}(\sum\limits_{t=1}^{T}\mathbf {\hat{y}}_{t}^{\mathrm{H}}\varepsilon(\mathbf{x})^{-1}\mathbf{\hat{y}}_{t})=-\beta^2  \sum_{t=1}^{T} \mathbf {\hat{y}}_{t}^{\mathrm{H}} \varepsilon(\mathbf{x})^{-1} \mathbf z_{k} \mathbf z_{k}^{\mathrm{H}} \varepsilon(\mathbf{x})^{-1}\mathbf{\hat{y}}_{t}.
	\end{aligned}
\end{align}

At the next step, $\bigtriangledown_{ x_k^{r}}(T\ln\det(\varepsilon(\mathbf{x})))$ is calculated as:
% Using the chain rule of the derivative formula, we can derive:
%$\bigtriangledown_{x_l^{\ell}}M\phi(\boldsymbol x^{\ell})$
\begin{align}\label{Eq: dTlog}
		\bigtriangledown_{ x_k^{r}}(T\ln\det(\varepsilon(\mathbf{x})))&=T\bigtriangledown_{ x_k^{r}}(\ln\det(\varepsilon(\mathbf{x})))\nonumber\\ &=\frac{T}{\det(\varepsilon(\mathbf{x}))}\bigtriangledown_{ x_k^{r}}(\det(\varepsilon(\mathbf{x}))).
\end{align}

Applying the matrix determinant derivative lemma~\cite{petersen2008matrix}, we obtain:
\begin{eqnarray}\label{Eq: ddet}
	\begin{array}{l}
		\bigtriangledown_{ x_k^{r}}(\det(\varepsilon(\mathbf{x})))
		=\det(\varepsilon(\mathbf{x})){\rm{tr}}(\varepsilon(\mathbf{x})^{-1}\bigtriangledown_{ x_k^{r}}(\varepsilon(\mathbf{x}))).
	\end{array}
\end{eqnarray}

From Eq. (\ref{Eq: dTlog}) and Eq. (\ref{Eq: ddet}), we have
\begin{eqnarray}\label{CCC}
	\begin{array}{l}
		\bigtriangledown_{ x_k^{r}}(T\ln\det(\varepsilon(\mathbf{x})))=T\cdot {\rm{tr}}{(\varepsilon(\mathbf{x})^{-1}\beta^2 \mathbf z_{k} \mathbf z_{k}^{\mathrm{H}})}.
	\end{array}
\end{eqnarray}

Finally, we have:
%$\bigtriangledown_{x_{l}} logf(\{\boldsymbol y_{m}\}|\boldsymbol x)$
\begin{align}\label{Eq: dlngx}
		\bigtriangledown_{ x_k^{r}}\ln g( \mathbf x^{r})=&\beta^2  \sum_{t=1}^{T} {\mathbf{\hat{y}}_{t}}^{\mathrm{H}} \varepsilon(\mathbf{x})^{-1} \mathbf z_{k} \mathbf z_{k}^{\mathrm{H}} \varepsilon(\mathbf{x})^{-1}\mathbf {\hat{y}}_{t}\nonumber\\&-T\cdot {\rm{tr}}{(\varepsilon(\mathbf{x})^{-1}\beta^2 {\mathbf z_{k}} {\mathbf z_{k}}^{\mathrm{H}})}.
\end{align}

\begin{figure*}[t]
%\vskip 0.2in
\begin{center}
\centerline{\includegraphics[width=1\linewidth]{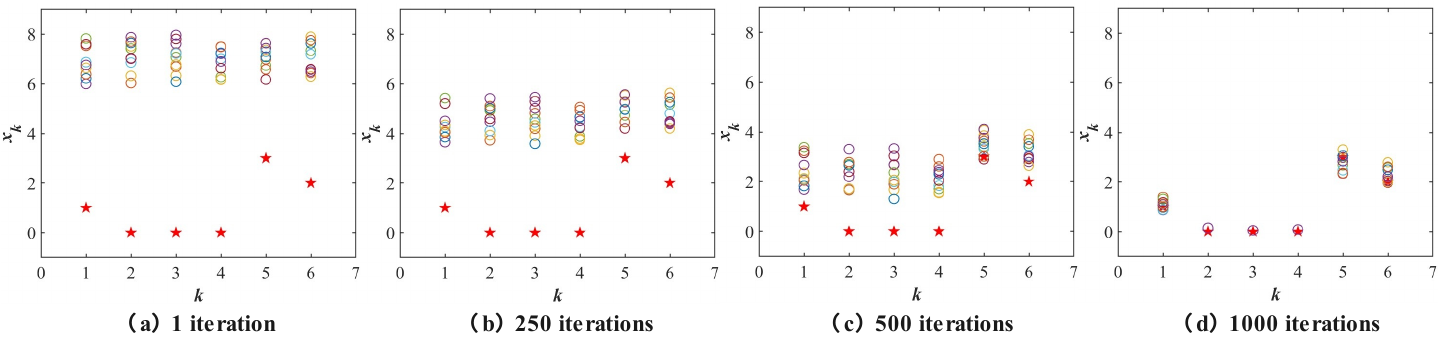}}
\caption{Updating process of particles. The circles represent particles and the pentagrams represent ground truth. $k$ is the index of the preamble and $x_k$ is the number of IoT devices choosing $k$-th preamble. $K=6$, $L=6$, $M=6$, $T=20$, and $\rm{SNR}=16$ dB.}
%\vskip -0.5in
\label{fig: iter}
\end{center}
\end{figure*}

After computing $\bigtriangledown_{ x_k^{r}}\ln g( \mathbf x^{r})$, we update the gradient according to Eq.~(\ref{Eq:gradient}).
Next, the history gradients are accumulated. Then, the accumulated gradient is normalized as follows:
\begin{equation}\label{Eq:accumulate the history gradients}
 {\omega}(\mathbf x_u^{r})\leftarrow \frac{{\omega}(\mathbf x_u^{r})}{\epsilon+\sqrt{\mathbf{q}_r}},
 \end{equation}
where $\epsilon$ is a constant and $\sqrt{(\cdot)}$ stands for element-wise square root.
These operations address the issue of the learning rate continually decreasing compared to the AdaGrad method~\cite{ruder2016overview}. Moreover, we employ weight decay~\cite{loshchilov2018fixing} and momentum strategy~\cite{atici2022normalized} to optimize the gradient ${\omega}(\mathbf x_u^{r})$ as shown in Algorithm \ref{alg: optimized SVGD}. Finally, the particles are updated according to Eq. (\ref{Eq: iter}).
After numerous iterations, the particles $\{\mathbf {\hat{x}}_{u}\}_{u=1}^n$ sampled by the blind NSVGD algorithm are employed to estimate the number of IoT devices choosing each preamble. For example, as shown in Fig. \ref{fig: iter}, we have six active IoT devices and six preambles in the cell. At first, ten particles are initialized using random numbers. As the number of iterations increases, the particles gradually approach the ground truth. After 1000 iterations, the particles converge to the ground truth.

It is observed that the blind NSVGD algorithm performs preamble detection without using the information of noise power and the number of active IoT devices compared with the NSVGD detector. This is because we apply the BMHT layer to denoise the complex signals before detecting preambles, which mitigates the vanishing gradients problem in the SVGD detector discussed in section~\ref{sec: NSVGD}. Therefore, we do not need to introduce any bias correction term to provide extra information to update the particles. 
The process of the blind NSVGD-based detector is summarized in Algorithm \ref{alg: optimized SVGD}.

\begin{algorithm}[t]
	\caption{Blind NSVGD-Based Detector}%算法名字
        \label{alg: optimized SVGD}
	\LinesNumbered %要求显示行号
	\KwIn{
 The received signals $\{\mathbf{y}_t^c\}_{t=1}^{T}$,
 a target function $g(\mathbf x)$ and  the initial particles $\{\mathbf x_{u}^0\}_{u=1}^n$}%输入参数
	\KwOut{A set of particles $\{\hat{\mathbf x}_{u}\}_{u=1}^n$ that approximates the target function}%输出	
    \tcc{Denoising stage}
    Divide $\{\mathbf{y}_t^c\}_{t=1}^{T}$ into $S$-length short-time windows. 

    \For{block i}{
    
    Perform the MHT using Eq. (\ref{Eq: MHT}).

    Remove the noise using Eq. (\ref{Eq: scaling}) and Eq (\ref{Eq: soft}).

    Perform the IMHT using Eq. (\ref{Eq: IMHT}).
    }
    \tcc{Preamble detection stage}
	\For{iteration $r$}{
        
        Compute $\bigtriangledown_{ x_k^{r}}\ln g( \mathbf x^{r})$ using Eq. (\ref{Eq: dlngx}).
        
        Update ${\omega}(\mathbf x_u^{r})$ according to Eq. (\ref{Eq:gradient}). 
        
        \tcc{Accumulate history gradients}
        \If {$\varrho \neq 0$}               
        {
            \uIf {$r > 1$} {
                ${\mathbf q}_r\leftarrow \varrho {\mathbf q}_{r-1}+(1-\varrho){\omega}^2(\mathbf x_u^{r})$
            } \Else {${\mathbf q}_r\leftarrow {\omega}^2(\mathbf x_u^{r})$} 
        }                

        Calculate ${\omega}(\mathbf x_u^{r})$ using Eq. (\ref{Eq:accumulate the history gradients}).
        
        \tcc{Weight Decay}
        \If{$\gamma \neq 0$}
        {
                ${\omega}(\mathbf x_u^{r})\leftarrow {\omega}(\mathbf x_u^{r})-\gamma \mathbf x_u^{r} $
        }

        \tcc{Gradient with Momentum}
        \If {$\alpha \neq 0$} 
        {
            \uIf {$r > 1$} {
                ${\mathbf b}_r\leftarrow \alpha {\mathbf b}_{r-1}+(1-\alpha){\omega}(\mathbf x_u^{r})$
            } \Else {${\mathbf b}_r\leftarrow {\omega}(\mathbf x_u^{r})$} 
            $ {\omega}(\mathbf x_u^{r})\leftarrow {\mathbf b}_r$
        }                       

        %\tcc{Gradient with Momentum}
		Compute $\mathbf x_u^{r+1}$ using Eq. (\ref{Eq: iter}).
	}
\end{algorithm}

\section{Experimental Results}\label{sec: simulation}

In the experimental section, we present the simulation results to explore the effectiveness of the algorithms proposed in our study. Specifically, to validate the denoising performance of the block MHT layer, other transform domain layers are selected as comparison models such as the DCT layer~\cite{pan2023real}, block wavelet transform perceptron (BWTP) layer~\cite{pan2023orthogonal} and Hadamard Transform Perceptron (HTP) layer~\cite{pan2023hybrid}. In the HTP layer, we set the number of channels as 3.
Besides, the convolutional neural network (CNN)~\cite{o2015introduction} and sparse autoencoder (AE)~\cite{ng2011sparse} are also utilized as comparison benchmarks because the convolutional and linear layers are widely employed for the denoising tasks. Additionally, to verify the preamble detection performance of the blind NSVGD-based detector, we choose other approximate inference-based detectors as comparison models, such as the MCMC detector~\cite{choi2018mcmc}, SVGD detector~\cite{zhu2024stein} and NSVGD detector~\cite{zhu2024stein}. 

In the experiments, we mainly consider a dense IoT device scenario by setting $K=M$ because our target is to detect preamble collision. 
Besides, since the noise power $\delta$ and the number of active users $M$ are unknown in the practical communication scenarios, all detectors perform preamble detection without using the information of $\delta$ and $M$.
Additionally, we consider non-orthogonal preambles. The elements of $\mathbf{Z}$ are independent CSCG random variables with zero-mean and variance $\frac{1}{L}$, \textit{i.e.}, $[{\rm {\mathbf{Z}}}]_{i,j}\sim \mathcal{CN}(0, \frac{1}{L})$. 
For all SVGD-based detectors, the particles $\{\mathbf {x}_{u}\}_{u=1}^n$ are initialized using a uniform distribution on $[1, 1.1]$. Then, the parameters $\beta$, $\lambda$, ${\kappa}$, $\rho$, $\epsilon$, and $\gamma$ are chosen as 1, 0.01, 0.001, 0.5, 1, and 0.1, respectively. 
Suppose $N$ denotes the number of iterations required to achieve stable particles. The sample mean of the particles is computed as follows: 
\begin{equation}
\bar{\mathbf x}=\frac{1}{n}\sum_{u=1}^{n}\hat{\mathbf x}_{u}=[\bar{x}_0,\dots,\bar{x}_k,\dots,\bar{x}_K].
\end{equation}
Then, $\tilde{x}_k$ is estimated by the rounded sample mean, $\textit{i.e.}$, $\tilde x_k=\lfloor\bar{x}_k\rceil\in\{0,\dots,M\}$, where $\lfloor x\rceil$ is the nearest integer of $x$.

\subsection{Performance metrics}

To evaluate the denoising performance of different methods, the percent root mean square difference (PRD) and root mean square (RMS) are employed. RMS measures the distinctions between the clean signal and the denoised signal.
\begin{equation}
{\rm{RMS}}=\sqrt{\frac{1}{w}\sum_{i=0}^{w-1}\left(s_i^c-s_i^d\right)^2},
\end{equation}
where $w$ is the length of the signal. $s_i^c$ is the clean signal. $s_i^d$ is the denoised signal. The lower the RMS is, the better denoising performance the approach has.

%\subsection{Percent root mean square difference (PRD)}
PRD evaluates the distortion of the denoised signal: 
\begin{equation}
{\rm{PRD}}=\sqrt{\frac{\sum_{i=0}^{w-1}\left(s_i^c-s_i^d\right)^2}{\sum_{i=0}^{w-1}\left(s_i^c\right)^2}}.
\label{Eq:prd}
\end{equation}
 The lower the PRD is, the better the model is.

Moreover, we utilize two performance metrics to evaluate the accuracy of the proposed preamble detection algorithm. Firstly, we consider the mean squared error (MSE). The MSE measures the difference between the true values and the estimated values. It is defined as follows:
	\begin{eqnarray}\label{MSE}
			{\rm{MSE}}(x_k)=\mathbb{E}(x_k-\tilde{x}_k)^2.
	\end{eqnarray}	
The lower the MSE is, the higher the accuracy of the preamble detection is. 

Another performance metric is the probability of activity detection error, which reflects the average proportion of incorrectly estimated preambles. It is defined as:
	\begin{eqnarray}\label{KKK}
		\begin{array}{l}
			{\rm{P_{ADE}}}={\rm {Pr}}(x_k\neq \tilde{x}_k).
		\end{array}
	\end{eqnarray}
The lower the $\rm{P_{ADE}}$ is, the higher the accuracy of the preamble detection is. 

Additionally, $N_p$ represents the number of trainable parameters. The Multiply–Accumulates (MACs) are employed to evaluate the computation complexity. One MAC represents one addition and one multiplication.

\subsection{Denoising Experiment}

In the denoising experiment, we generate a dataset using the signal reception model defined in Eq. (\ref{Eq:yph}). 
% Specifically, we use the received signal $\mathbf y\in\mathbb{C}^{L}$ on a single antenna as one data sample. Since neural networks cannot be trained with complex numbers, we concatenate the real and imaginary parts of $\mathbf y$ as $\mathbf y_c\in\mathbb{R}^{2L}$.
Specifically, we generate 1750 noisy samples as inputs and corresponding clean samples as labels by setting $K=20$, $L=10$, $M=20$ and $\rm{SNR}=10$ dB. 
These data samples are employed as training datasets. Furthermore, to test the generalization capability of denoising models, another 1750 sample pairs are generated as testing datasets by changing $\rm{SNR}$ to 6 dB. Additionally, during the training stage, we use the AdamW optimizer~\cite{loshchilov2017decoupled}. The learning rate and batch size are set as 0.001 and 128 respectively.

Table \ref{tab: Denoising Experimental Results} presents the denoising results of different models on the testing datasets. Compared with CNN, the BMHT layer reduces RMS from 40.55 to 34.02 (16.10\%) and PRD from 40.50 to 33.98 (16.10\%). Additionally, $N_p$ decreases from 824 to 16 (98.06\%) and MACs decrease from 840 to 216 (74.29\%). Although CNN and sparse AE apply more trainable parameters than the BMHT layer, the BMHT layer still achieves a better denoising performance than CNN and sparse AE. It indicates too many training parameters can result in overfitting of the model in the case of RA. 
% Moreover, the sparse AE and the BMHT layer both use the penalty term to reduce noise. However, the BMHT layer outperforms the sparse AE in terms of RMS, PRD and MACs.  
Additionally, the BMHT layer is superior to other familiar transform-based models such as the DCT layer, BWTP layer and HTP layer because of a lower RMS and PRD. The reason is that the KL divergence-based sparsity penalty is applied in the BMHT layer to keep latent space sparse, which helps remove noise in the transform domain. Besides, in comparison to the HTP layer, no convolutional layers are implemented in the BMHT layer, which reduces the computation costs and number of parameters.

The ablation experiments are used to validate the function of each module in the BMHT layer. As shown in Table \ref{tab: Ablation study}, when the scaling layer is removed, the denoising performance degrades as the RMS increases from 34.02 to 35.32 (3.82\%) and PRD from 33.98 to 35.27 (3.80\%). 
It means the scaling layer can reduce the noise by assigning appropriate emphasis to frequency domain components. 
Additionally, when the sparsity penalty and soft thresholding layer are not present in the BMHT layer, the RMS and PRD both increase. It shows that the penalty term and soft threshold are beneficial to retaining important features and eliminating redundant information such as noise. Furthermore, it is observed that the RMS and PRD increase when the MHT is replaced with the standard Hadamard transform (HT). Therefore, it is experimentally shown that the MHT has a better ability to separate the noise from the signals than the standard Hadamard transform.  

\begin{table}[t]
%\begin{center}
\begin{small}
\begin{sc}
\caption{Denoising Experimental Results.}
    \label{tab: Denoising Experimental Results}
%\vskip 0.02in
    \centering
    \begin{tabular}{lcccc}
    \toprule
       Method&$N_p$&MACs&RMS (\%)&PRD (\%)\\
        \midrule
        CNN~\cite{o2015introduction} &824&840&40.55&40.50\\
%        AE-DCST &272,474&41.32&27.36\\
        DCT Layer~\cite{pan2023real}&40 &820&34.16&34.11\\
        Sparse AE~\cite{ng2011sparse}&676 &640 &53.77&53.70\\
        BWTP Layer~\cite{pan2023orthogonal} &51&550&34.17&34.12\\
        HTP Layer~\cite{pan2023hybrid} &195&1216 &34.16&34.12\\
        
        \bf{BMHT LAYER} &\bf{16} &\bf{216} &\bf{34.02}&\bf{33.98}\\

        \bottomrule
\end{tabular}
\vskip -0.1in
\end{sc}
\end{small}
%\end{center}
\end{table}

\begin{table}[t]
%\begin{center}
\begin{small}
\begin{sc}
\caption{Ablation study.}
    \label{tab: Ablation study}
%\vskip 0.02in
    \centering
    \begin{tabular}{lcccc}
    \toprule
       Method&$N_p$&MACs&RMS (\%)&PRD (\%)\\
        \midrule
        No penalty &16&216&34.16&34.12\\
%        AE-DCST &272,474&41.32&27.36\\
        No scaling&\bf{8} &\bf{192}&35.32&35.27\\
        No threshold&8 &216 &34.16&34.12\\
        Standard HT&16 &216 &35.05&35.00\\
        \bf{BMHT LAYER} &{16} &{216} &\bf{34.02}&\bf{33.98}\\
        \bottomrule
\end{tabular}
\vskip -0.1in
\end{sc}
\end{small}
%\end{center}
\end{table}

% \begin{table}[t]
% %\begin{center}
% \begin{small}
% \begin{sc}
% \caption{Ablation study.}
%     \label{tab: Ablation study}
% %\vskip 0.02in
%     \centering
%     \begin{tabular}{lcccc}
%     \toprule
%        Method&RMSE (\%)&PRD (\%)\\
%         \midrule
%         No penalty &27.47&27.43\\
% %        AE-DCST &272,474&41.32&27.36\\
%         No scaling&28.04&27.99\\
%         No threshold&27.49&27.44\\
%         Standard HT&27.85&27.81\\
%         \bf{SMHT LAYER} &\bf{27.38}&\bf{27.34}\\
%         \bottomrule
% \end{tabular}
% \vskip -0.1in
% \end{sc}
% \end{small}
% %\end{center}
% \end{table}

\subsection{Preamble Detection Experiments}

In the preamble detection experiments, to test the generalization ability 
of the proposed model, we use the fixed BMHT layer, which is trained when $K=20$, $L=10$, $M=20$ and $\rm{SNR}=10$ dB. It means that even if $K$, $M$ and $\rm{SNR}$ change in the environment, we do not train the BMHT layer again.

\begin{figure}[t]
%\vskip 0.2in
\begin{center}
\centerline{\includegraphics[width=0.75\linewidth]{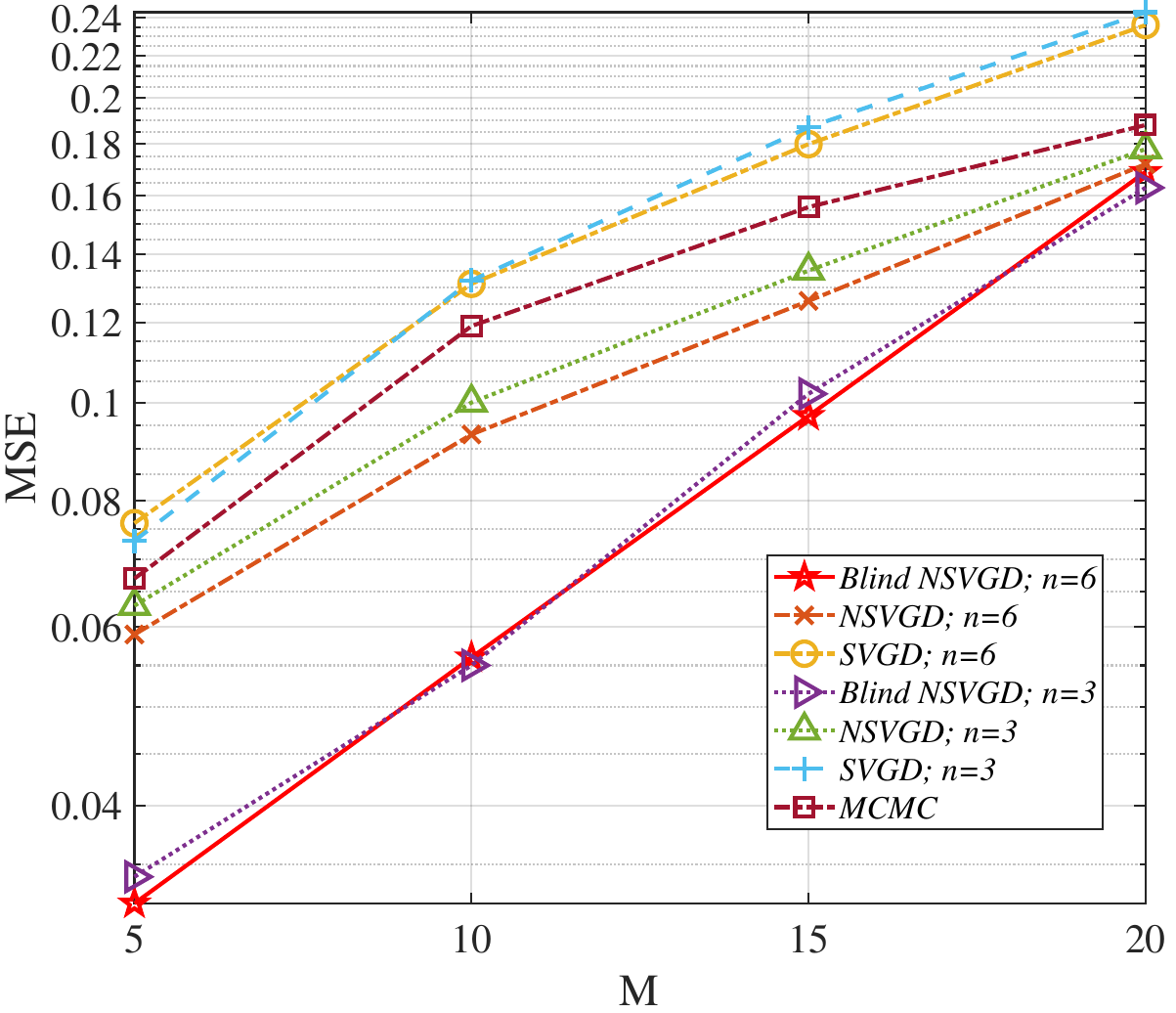}}
\caption{MSE for different numbers of active IoT devices when $K=20$, $L=10$, $T=35$ and $\rm{SNR}=8$ dB.}
%\vskip -0.5in
\label{fig: M_MSE}
\end{center}
\end{figure}

\begin{figure}[t]
%\vskip 0.2in
\begin{center}
\centerline{\includegraphics[width=0.75\linewidth]{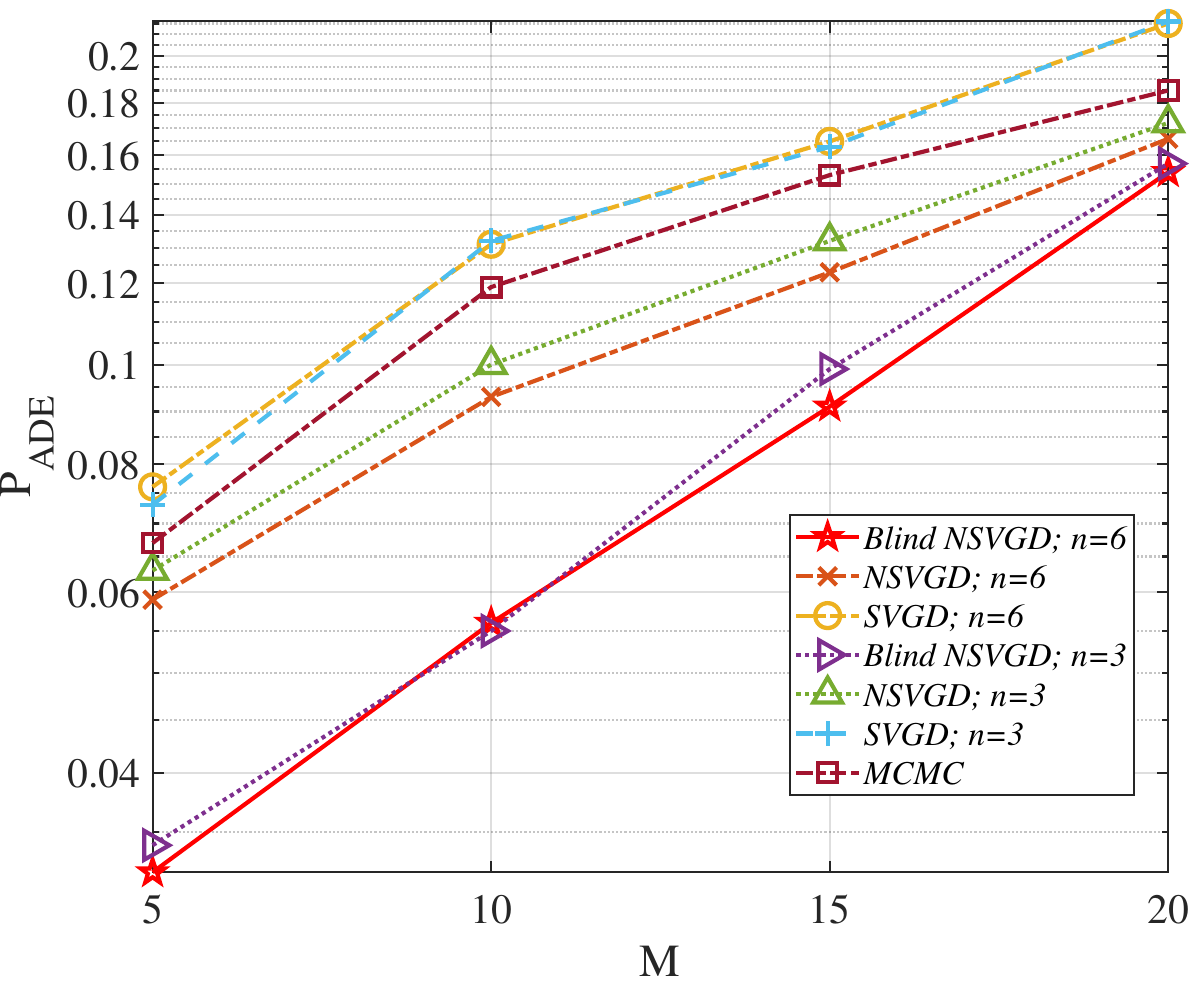}}
\caption{$\rm{P_{ADE}}$ for different numbers of active IoT devices when $K=20$, $L=10$, $T=35$ and $\rm{SNR}=8$ dB. }
%\vskip -0.5in
\label{fig: M_Pade}
\end{center}
\end{figure}

\begin{figure}[t]
%\vskip 0.2in
\begin{center}
\centerline{\includegraphics[width=0.76\linewidth]{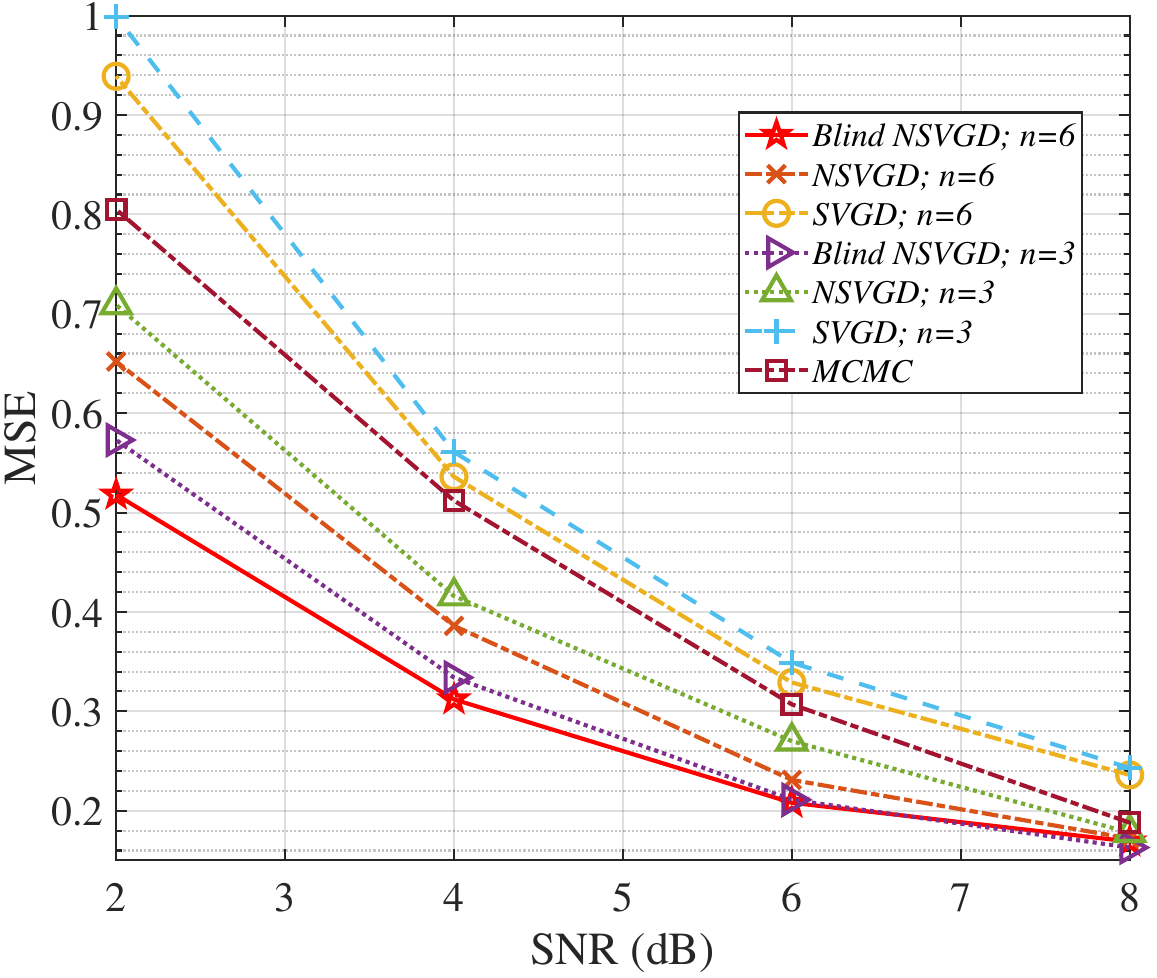}}
\caption{MSE for different SNR when $K=20$, $L=10$, $M=20$ and $T=35$.}
%\vskip -0.5in
\label{fig: SNR_MSE}
\end{center}
\end{figure}

\begin{figure}[t]
%\vskip 0.2in
\begin{center}
\centerline{\includegraphics[width=0.76\linewidth]{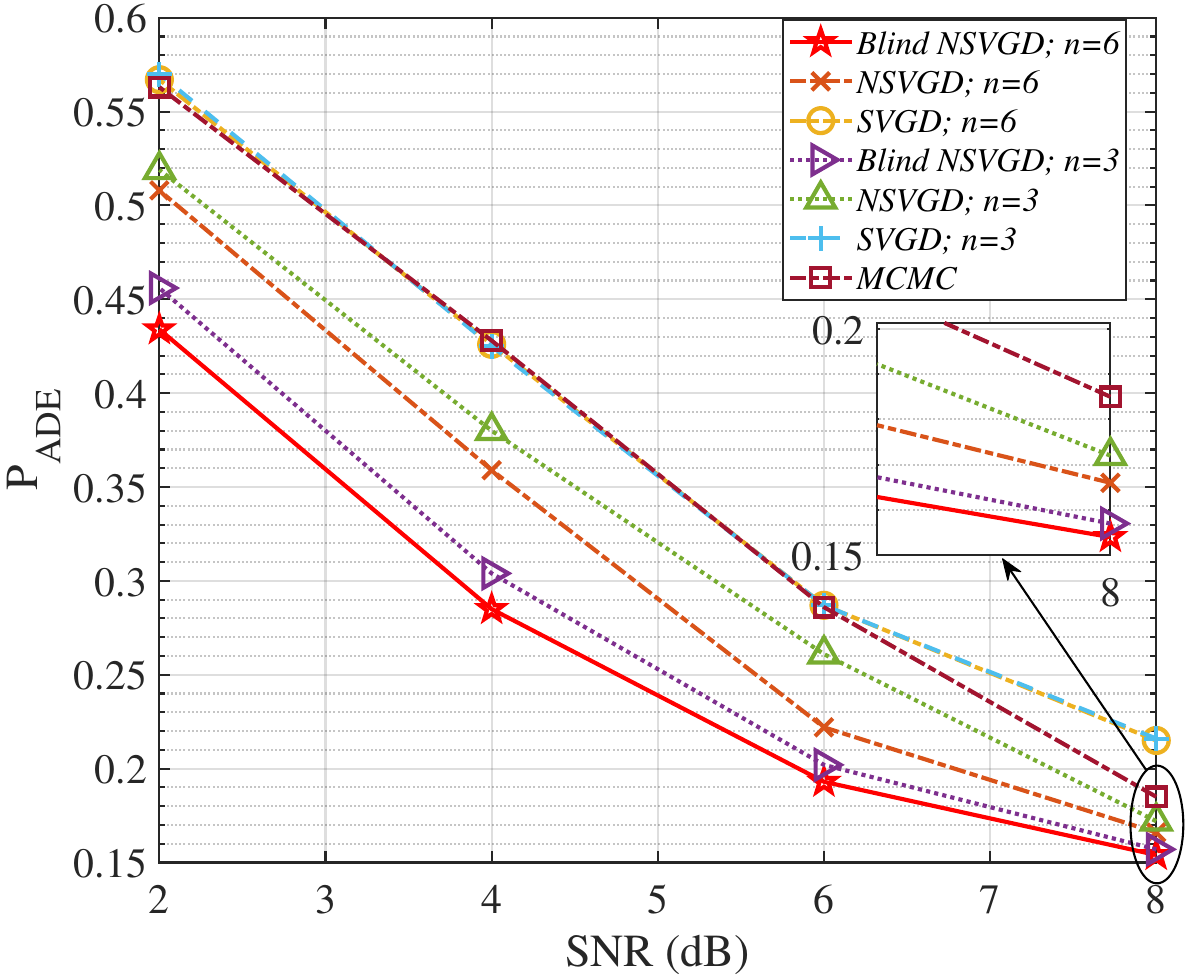}}
\caption{$\rm{P_{ADE}}$ for different SNR when $K=20$, $L=10$, $M=20$ and $T=35$.}
%\vskip -0.5in
\label{fig: SNR_pade}
\end{center}
\end{figure}

% \begin{figure}[t]
% %\vskip 0.2in
% \begin{center}
% \centerline{\includegraphics[width=0.7\linewidth]{figure/T_MSE.jpg}}
% \caption{MSE for different numbers of $T$ when $K=20$, $L=10$, $M=20$ and $\rm{SNR}=8$ dB.}
% %\vskip -0.5in
% \label{fig: T_MSE}
% \end{center}
% \end{figure}

% \begin{figure}[t]
% %\vskip 0.2in
% \begin{center}
% \centerline{\includegraphics[width=0.7\linewidth]{figure/T_pade.jpg}}
% \caption{$\rm{P_{ADE}}$ for different numbers of $T$ when $K=20$, $L=10$, $M=20$ and $\rm{SNR}=8$ dB.}
% %\vskip -0.5in
% \label{fig: T_pade}
% \end{center}
% \end{figure}

\begin{figure}[t]
%\vskip 0.2in
\begin{center}
\centerline{\includegraphics[width=0.76\linewidth]{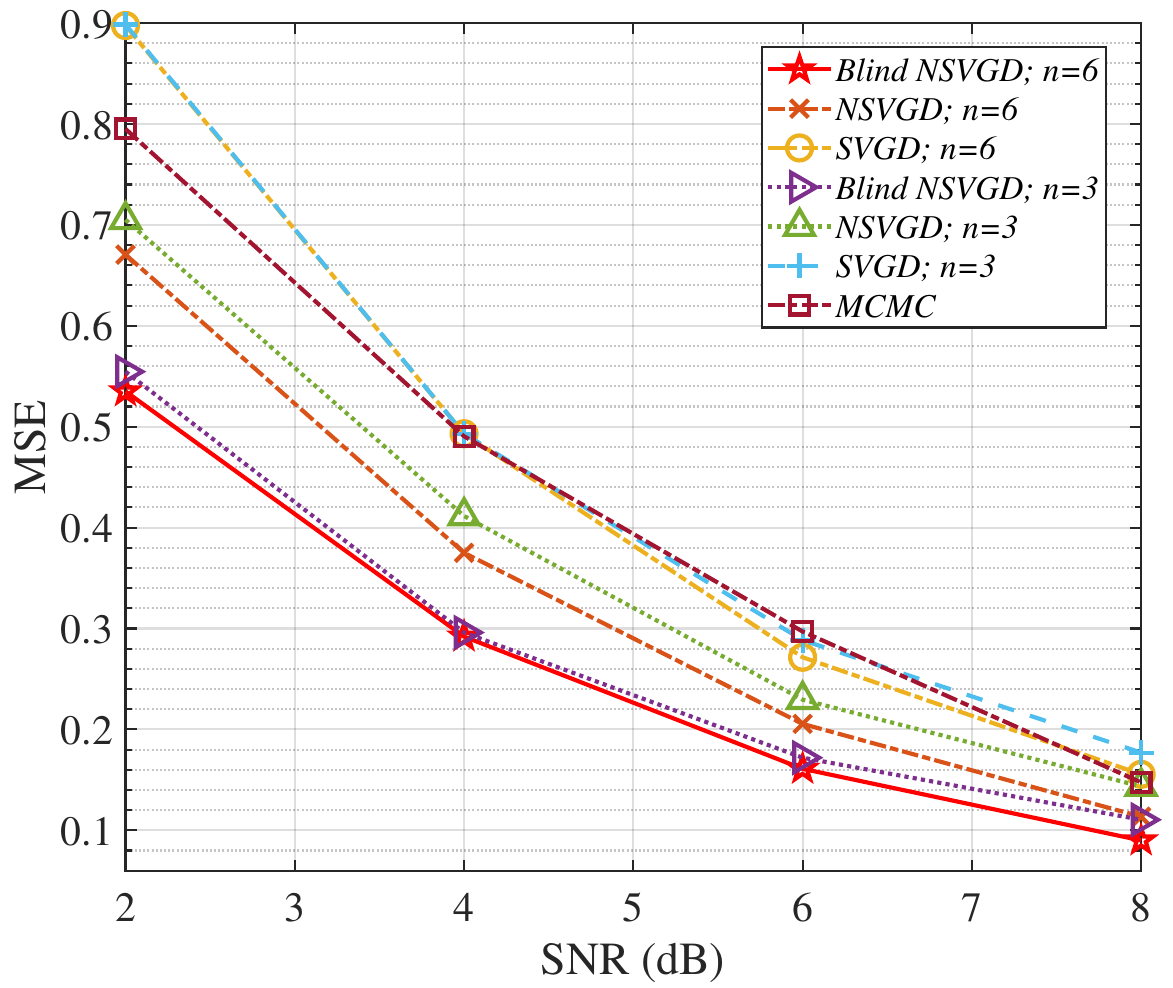}}
\caption{MSE for different SNR when $K=20$, $L=10$ and $T=35$ in a network with dynamic delay-sensitive devices.}
%\vskip -0.5in
\label{fig: SNR_MSE_dy}
\end{center}
\end{figure}

\begin{figure}[t]
%\vskip 0.2in
\begin{center}
\centerline{\includegraphics[width=0.76\linewidth]{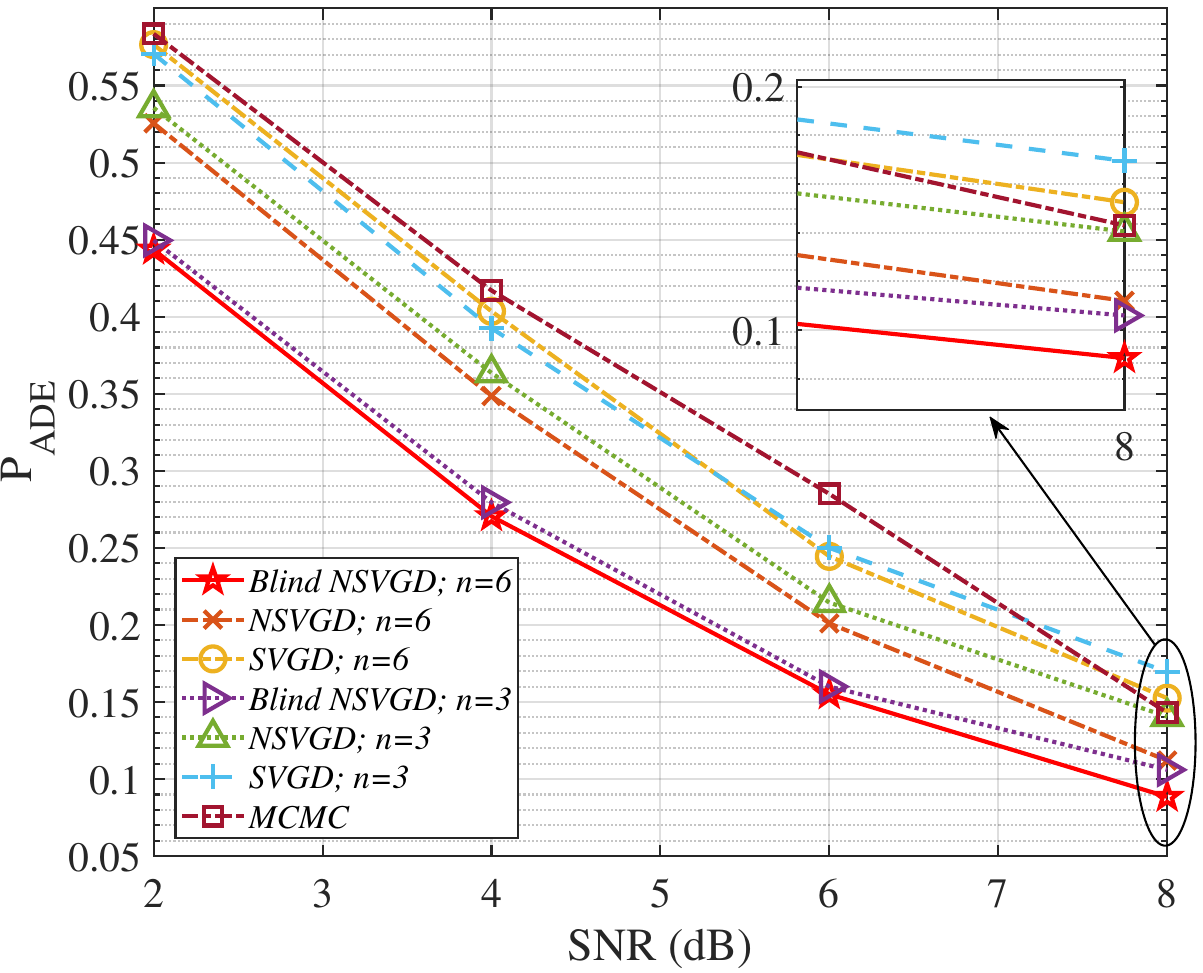}}
\caption{$\rm{P_{ADE}}$ for different SNR when $K=20$, $L=10$ and $T=35$ in a network with dynamic delay-sensitive devices.}
%\vskip -0.5in
\label{fig: SNR_pade_dy}
\end{center}
\end{figure}

\begin{figure}[t]
%\vskip 0.2in
\begin{center}
\centerline{\includegraphics[width=0.76\linewidth]{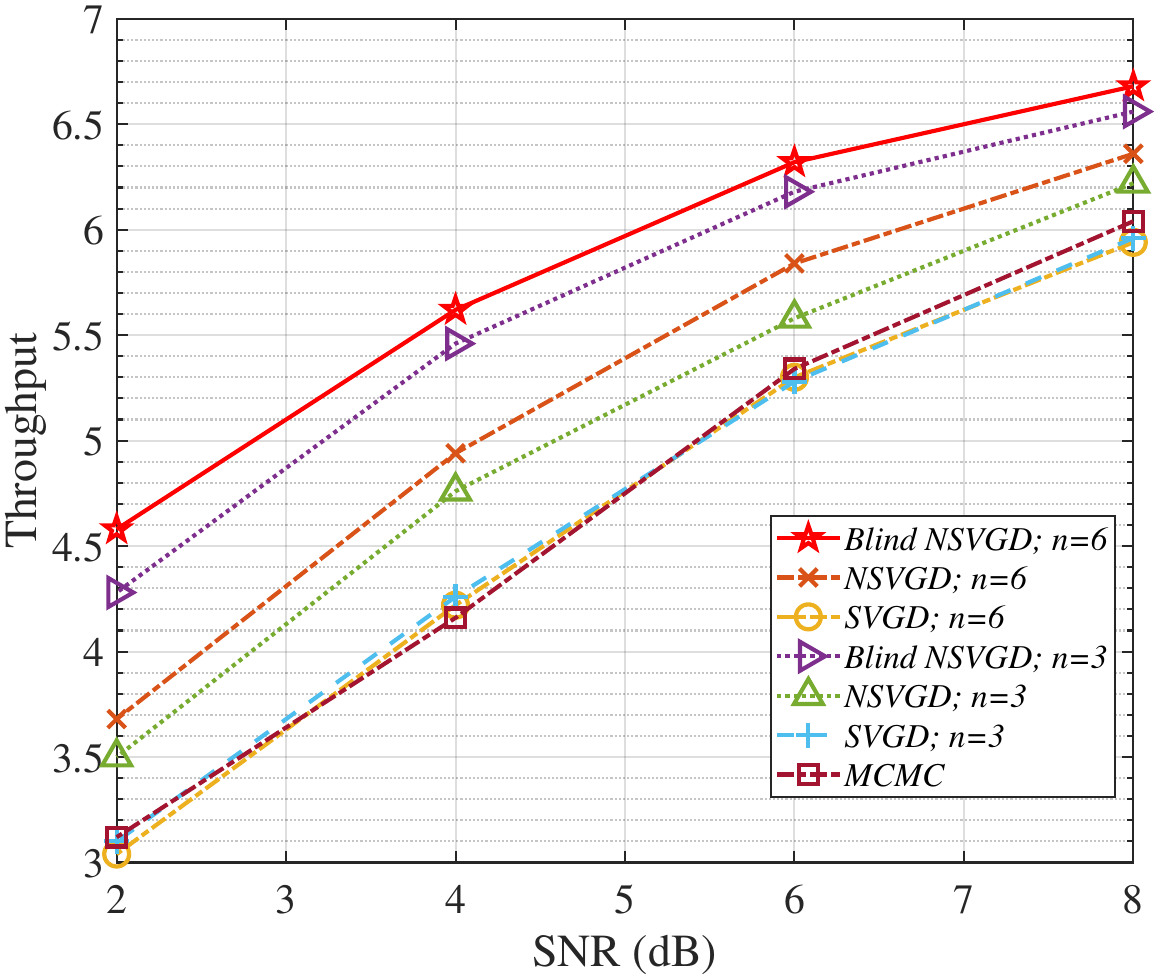}}
\caption{Throughput for different SNR when $K=20$, $L=10$, $M=20$ and $T=35$.}
%\vskip -0.5in
\label{fig: TH}
\end{center}
\end{figure}

% Fig. \ref{fig: T_MSE} and Fig. \ref{fig: T_pade} show the effect of different numbers of antennas on the MSE and $\rm{P_{ADE}}$. With the number of antennas increasing, both the MSE and $\rm{P_{ADE}}$ decrease, leading to a continuous reduction in preamble detection error. This is because more antennas introduce diversity in receiving signals, which alleviates noise and fading. Additionally, the performance of the blind NSVGD-based detector remains superior to that of the MCMC, NSVGD and SVGD detectors. 
% It indicates that the proposed detector has better robustness than other baselines. 

In Fig. \ref{fig: M_MSE} and Fig. \ref{fig: M_Pade}, we show the performances of the MCMC, SVGD, NSVGD and blind NSVGD-based detectors for different numbers of active IoT devices. When the number of active IoT devices increases, the MSE and $\rm{P_{ADE}}$ both increase. The reason is that the number of preamble collisions increases when more active IoT devices try to access the network simultaneously. Severe preamble collisions bring much interference, which makes it difficult for approximate inference-based methods to perform preamble detection. Additionally, the blind NSVGD-based detector still outperforms other detectors under different $M$, which indicates the proposed method can achieve good results in both sparse and dense IoT device scenarios. 
When $M=5$, the $\rm{P_{ADE}}$ of blind NSVGD-based detector is only around 0.03. It means that, on average, no more than 1 out of 20 preambles is detected incorrectly. Even in the case of severe collisions ($M=20$), only around 3 out of 20 preambles are detected incorrectly on average.
Besides, when the number of particles $n$ in the SVGD, NSVGD, and blind NSVGD-based detectors increases from 3 to 6, the preamble detection accuracy also improves. It implies that increasing $n$ helps reduce detection errors and enhance the robustness of detectors at different $M$.

Fig. \ref{fig: SNR_MSE} and Fig. \ref{fig: SNR_pade} show the preamble detection performance of four models at different SNR. As the SNR increases, the MSE and $\rm{P_{ADE}}$ of each model decrease, which indicates an improvement in detection accuracy. 
It is observed that the blind NSVGD-based detector surpasses the MCMC, SVGD and NSVGD detectors in terms of MSE and $\rm{P_{ADE}}$. 
Additionally, when SNR $= 8$ dB and $n=6$, the proposed method exhibits 1\% performance improvement over the NSVGD detector in terms of $\rm{P_{ADE}}$ as shown in Fig. \ref{fig: SNR_pade}. As SNR decreases, the performance gap between the proposed method and the NSVGD detector also widens. The reason is that the BMHT layer achieves good denoising performance under low SNR. Therefore, the blind NSVGD-based detector can extract more efficient gradient information to update the particles without prior knowledge of active IoT devices and noise power. 

To assess the robustness of the proposed detector in more heterogeneous or dynamic network environments, we assume that the system includes a subset of delay-sensitive devices. Therefore, the active devices can be divided into two categories: delay-sensitive and delay-tolerant. Delay-sensitive devices refer to those that have strict requirements regarding access latency. Moreover, the number of delay-sensitive and delay-tolerant devices changes dynamically across time slots. 
In the experiment, we assume $K=20$, $L=10$ and $T=35$. The number of delay-sensitive devices fluctuates between 3 and 7, while the number of delay-tolerant devices varies between 8 and 12.  
To reduce collision and access latency for delay-sensitive devices, 10 preambles are exclusively allocated to them, while the remaining preambles are assigned to delay-tolerant devices. Each active device randomly selects a preamble from the assigned pool. 
Other environment configurations remain consistent with those described in Section~\ref{sec: System}.
In the prior preamble detection experiment, the number of active users is fixed across different SNR environments. In contrast, the current experiment features a dynamic variation in the number of active users across different SNR.
Besides, the grouping strategy restricts the preamble selection range for two different types of active devices, reducing the risk of preamble collisions among delay-sensitive devices, especially under high load conditions.
Fig. \ref{fig: SNR_MSE_dy} and Fig. \ref{fig: SNR_pade_dy} illustrate the detection performance of various detectors across different SNR values. Compared to the NSVGD detector ($n$=6), the blind NSVGD-based detector achieves a reduction in MSE from 0.1135 to 0.0900, representing a 20.70\% improvement, and a reduction in $\rm{P_{ADE}}$ from 0.1120 to 0.0885, corresponding to an enhancement of 20.98\%, at an SNR of 8 dB. Moreover, the blind NSVGD-based detector ($n$=6) outperforms the SVGD and MCMC detectors due to the lowest MSE and $\rm{P_{ADE}}$. Therefore, the proposed detector exhibits strong robustness in a heterogeneous and dynamic network environment. This is attributed to the block MHT layer's ability to eliminate noise from the received signals. By reducing noise, the blind NSVGD algorithm can concentrate on the critical features of the data, avoiding distraction from irrelevant or random disturbances. As a result, it achieves more stable and reliable preamble detection outcomes compared to other state-of-the-art models.

Fig. \ref{fig: TH} compares the throughputs of different detectors. Throughput refers to the average count of successfully accessed preambles, excluding instances of collisions and false detections. As SNR decreases, the throughputs of different detectors degrade. The reason is that the noise influences the preamble detection accuracy of detectors. Moreover, when SNR is 6 dB, compared with the SVGD detector, the blind NSVGD-based detector achieves a significant performance improvement as it increases throughput from 5.30 to 6.32 (19.25\%). Additionally, the blind NSVGD-based detector performs better than the NSVGD and MCMC detectors. It indicates even when severe preamble collisions happen at different SNR, the proposed detector still can achieve higher accuracy in detecting the successfully accessed preambles. Besides, throughputs of SVGD, NSVGD, and blind NSVGD-based detectors rise with $n$ increasing.

\subsection{The analysis of computational complexity}
We further present an analysis of the computational complexity of the blind NSVGD-based detector, which is dominated by the number of multiplication operations as discussed in Section~\ref{sec: BSVGD}. For the block MHT layer, performing the MHT and IMHT calculations requires $O(TL)$ multiplications. The scaling layer contributes an additional $O(TL)$ multiplications. Moreover, the product involving $\text{sign}\left(\mathbf{\bar{y}}{t}\right)$ and $\left(|\mathbf{\bar{y}}{t}|-\mathbf{T}\right)_{+}$ can be realized using sign-bit operations~\cite{pan2023hybrid}, which means that the soft-thresholding step does not necessitate any multiplications. Thus, the computational complexity of the block MHT layer is $O(TL)$. For the blind NSVGD algorithm, The complexity of each iteration is primarily determined by the cost of computing Eq.~(\ref{Eq: dlngx}), which requires $O(nKTL^2)$ multiplications. Consequently, the complexity of the blind NSVGD algorithm is $O(nKTL^2)$. The computational complexity of the block MHT layer can be omitted compared to that of the Blind NSVGD algorithm. Finally, the overall complexity of the blind NSVGD-based detector is $O(nKTL^2)$. Additionally, the computation complexities of the MCMC detector, SVGD detector and NSVGD detector are $O(KL(L+T))$, $O(nKTL^2)$ and $O(nKTL^2)$ respectively~\cite{choi2018mcmc,zhu2023stein}. Hence, the SVGD detector, NSVGD detector, and blind NSVGD-based detector have similar computational complexity. When $n$ is large ($n\gg\frac{L+T}{TL}$), the blind NSVGD-based detector incurs greater computational complexity compared to the MCMC detector. 
% However, the MCMC detector has a lower detection accuracy than the blind NSVGD-based detector as shown in Figs~\ref{fig: M_Pade} and~\ref{fig: SNR_pade}.
Additionally,
given a fixed $L$ and $n$, the complexities of all detectors are linearly proportional to $K$. Besides, they are independent of the number of active IoT devices.

% The computation complexities of the MCMC detector, SVGD detector and NSVGD detector are $O(KL(L+T))$, $O(nKTL^2)$ and $O(nKTL^2)$ respectively~\cite{zhu2024stein,choi2018mcmc}. Additionally, compared with the NSVGD detector, the blind NSVGD-based detector introduces the extra computation cost of the BMHT layer. The computation complexity of the BMHT layer is $O(L)$. Therefore, it can be omitted compared with $O(nKTL^2)$. Finally, the overall complexity of the blind NSVGD-based detector is $O(nKTL^2)$. Hence, given a fixed $L$, the complexities of all detectors are linearly proportional to $K$. Besides, they are independent of the number of active IoT devices.  

\begin{figure}[t]
%\vskip 0.2in
\begin{center}
\centerline{\includegraphics[width=0.76\linewidth]{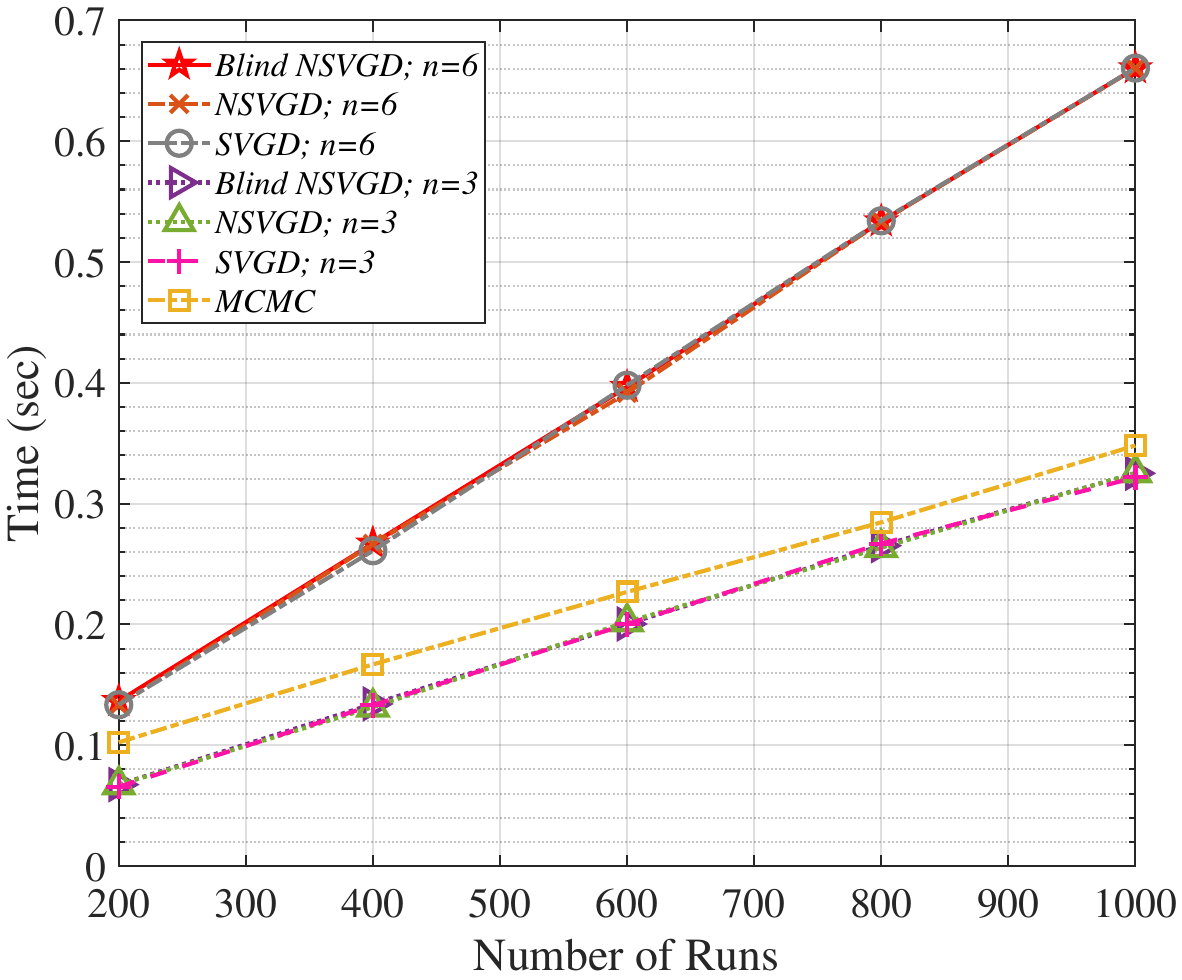}}
\caption{ Computation times for different numbers of runs when $K=20$, $L=10$, $M=20$, $T=1$ and $\rm{SNR}=8$ dB.}
%\vskip -0.5in
\label{fig: Time}
\end{center}
\end{figure}

Fig. \ref{fig: Time} compares the computation times of different detectors for different numbers of runs. Since the received signals can be computed in parallel on multiple antennas, we only consider the computation times on a single antenna, which reflects the latency of different detectors. 
The simulation is executed on the Intel Core i7-12700H CPU.
When $n=3$ (close to $\frac{L+T}{TL}$),  the runtime of the blind NSVGD-based detector is similar to that of the SVGD, NSVGD, and MCMC detectors. Nevertheless, the blind NSVGD-based detector achieves superior preamble detection accuracy as shown in Figs.~\ref{fig: M_MSE}-\ref{fig: TH}.
As $n$ increases to 6 (much larger than $\frac {L+T}{TL}$), the runtime of the blind NSVGD-based detector remains similar to that of the SVGD and NSVGD detectors but becomes higher than that of the MCMC detectors. Therefore, the blind NSVGD-based detector achieves a trade-off between the preamble detection accuracy and computation complexity when $n=3$.

% However, it exhibits a slightly shorter runtime compared to the MCMC detector. This is because the stochastic nature of sampling makes it challenging for the MCMC detector to converge, thereby requiring a greater number of iterations.
% When $n=3$, the run time of the blind NSVGD-based detector is similar to that of the MCMC detector.
% What is more, increasing $n$ by a factor of two leads to a twofold increase in the run times of the SVGD-based detectors, leading to a longer runtime than that of the MCMC detector.
In addition, the blind NSVGD-based detector has a runtime of less than 0.7 seconds, indicating that it introduces minimal latency to the RA process. Therefore, the blind NSVGD-based detector can achieve a balance between preamble detection accuracy and computation time.

\section{Conclusion}
In this paper, we propose an early preamble detection scheme based on the blind NSVGD-based detector at the first step of the GBRA scheme in cellular IoT to reduce resource wastage. The blind NSVGD-based detector consists of two modules: a BMHT layer and a blind NSVGD algorithm. At the receiver, the MHT is first conceived. It separates high-frequency components from the signal in the transform domain. After that, the BMHT layer is designed based on the MHT, trainable scaling layer and soft-thresholding layer, which removes the noise and alleviates the issue of vanishing gradients in the SVGD-based detectors. Finally, the derived blind NSVGD algorithm completes the preamble detection task without requiring unknown prior knowledge. The simulation results demonstrated the BMHT layer outperformes other denoising methods with a low computation cost. Additionally, aided by the BMHT layer, the proposed detector has a consistent performance improvement over the MCMC detector and other SVGD-based detectors in terms of detection accuracy and throughput.

% \section*{Acknowledgments}
% Xin Zhu was supported by the National Science Foundation (NSF) under grant 1934915 and NSF IDEAL 2217023.

% {\appendix[Proof of the Zonklar Equations]
% Use $\backslash${\tt{appendix}} if you have a single appendix:
% Do not use $\backslash${\tt{section}} anymore after $\backslash${\tt{appendix}}, only $\backslash${\tt{section*}}.
% If you have multiple appendixes use $\backslash${\tt{appendices}} then use $\backslash${\tt{section}} to start each appendix.
% You must declare a $\backslash${\tt{section}} before using any $\backslash${\tt{subsection}} or using $\backslash${\tt{label}} ($\backslash${\tt{appendices}} by itself
%  starts a section numbered zero.)}


% \section{References Section}
% You can use a bibliography generated by BibTeX as a .bbl file.
%  BibTeX documentation can be easily obtained at:
%  http://mirror.ctan.org/biblio/bibtex/contrib/doc/
%  The IEEEtran BibTeX style support page is:
%  http://www.michaelshell.org/tex/ieeetran/bibtex/
 
\bibliographystyle{IEEEbib}
\bibliography{reference}

\begin{IEEEbiography}[{\includegraphics[width=1in,height=1.25in,clip,keepaspectratio]{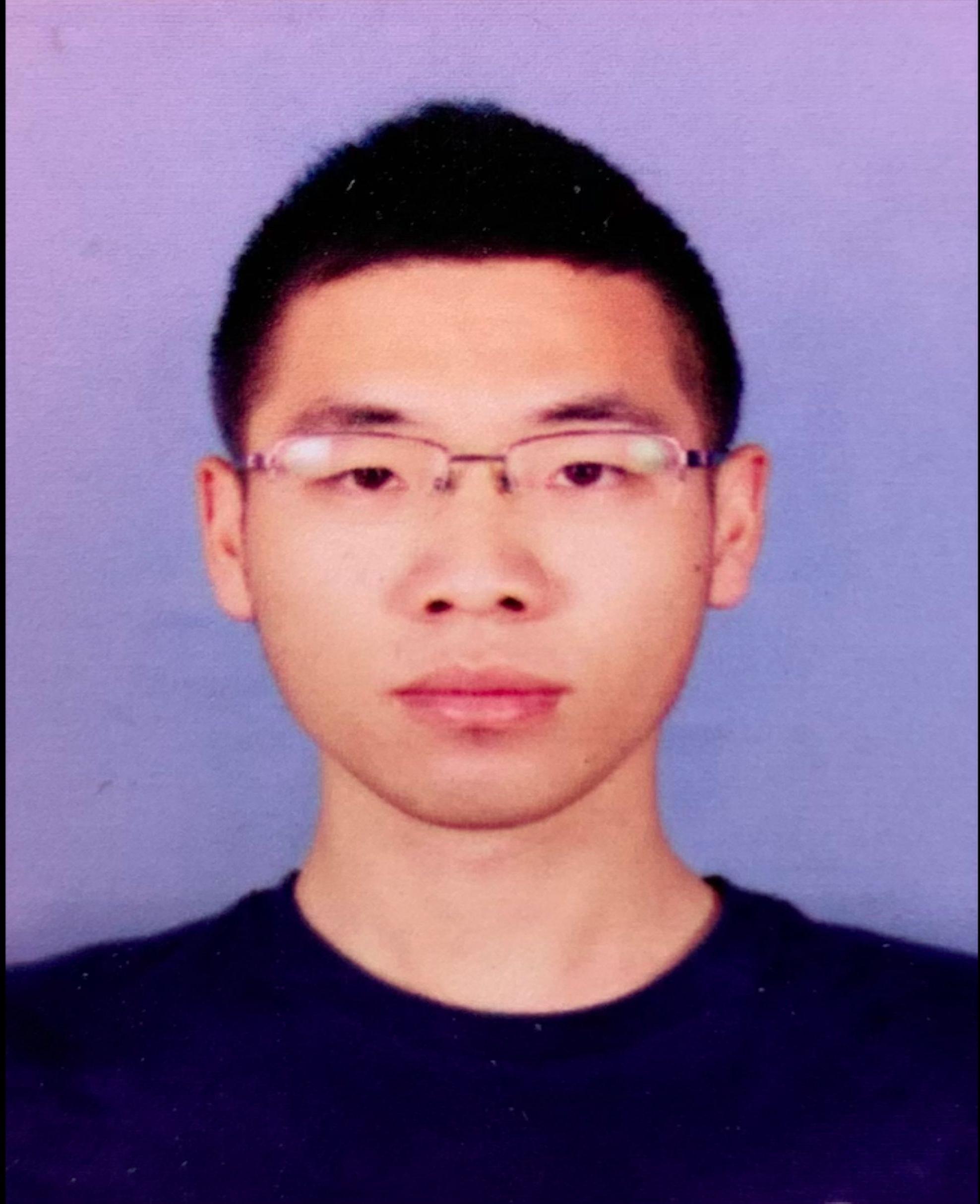}}]{X. Zhu}{\space}received the M.S. degree from the School of Telecommunications Engineering, Xidian University, Xi'an, China, in 2022. He is currently pursuing the Ph.D. degree in the Department of Electrical and Computer Engineering at the University of Illinois Chicago, Illinois, USA. His research interests include signal processing, machine learning, data compression, random access, and biomedical image analysis.
\end{IEEEbiography}

\begin{IEEEbiography}[{\includegraphics[width=1in,height=1.25in,clip,keepaspectratio]{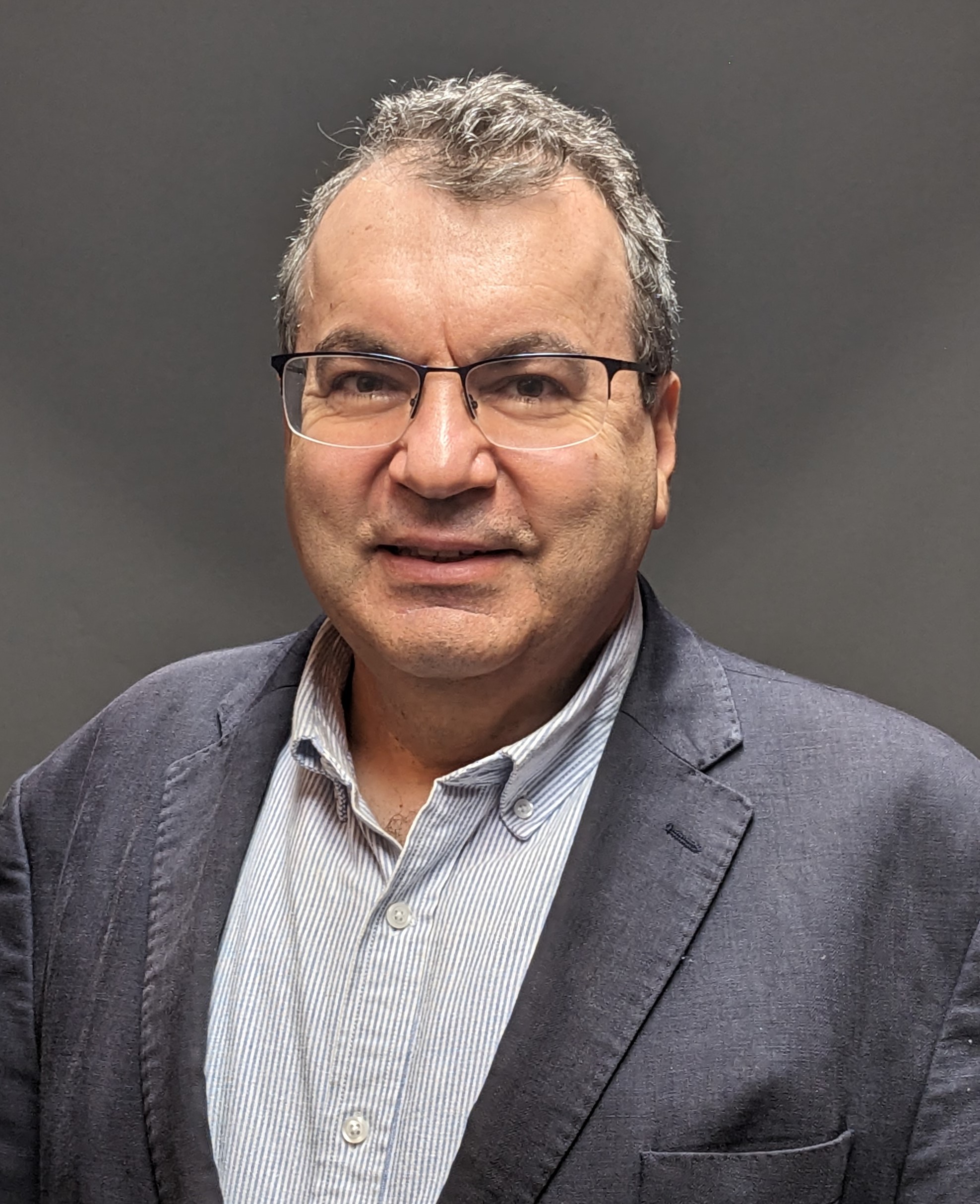}}]{A. E. Cetin}{\space}received his B.Sc. from METU, Ankara, Turkey, and Ph.D. in 1987 from the University of Pennsylvania, USA. He was an Assistant Prof. at the University of Toronto between 1987-1989 in Canada. He was a faculty member at Bilkent University from 1987 to 2017. He is currently a professor in the Department of Electrical and Computer Engineering at the University of Illinois Chicago (UIC). He also held visiting professor positions at Bellcore (1988), University of Minnesota (1996-1997), and UC San Diego (2016-2017). He has been carrying out research in the areas of theoretical and applied machine learning, signal, image, and video processing, biomedical signal processing, infrared and chemical sensor signal processing in Cyber-Physical Systems (CPS). His group introduced the concept of adaptive prediction and split vector quantization for Line Spectral Frequency representation. This concept was used in ITU speech coding standards including G.729, G.723.1, and GSM EFR.

He became a Fellow of IEEE for his contributions to signal and image recovery. He is the Editor-in-Chief of Signal, Image and Video Processing, Springer-Nature. He received a best paper award for his camera-based wildfire detection work at a conference organized by UNESCO and Cyprus Presidency of the European Union.  He is one of the co-founders of the multinational smart wide-angle OEM camera company Oncam-Grandeye (https://www.oncamgrandeye.com), UK. He served as the CEO/CTO of Grandeye, Turkey between 2003-2013. Oncam-Grandeye cameras won design and innovation awards in IFSEC, UK, and ISC WEST, Las Vegas, trade fairs.

\end{IEEEbiography}

\vfill

\end{document}